\documentclass[useAMS,usenatbib]{mnras}
\usepackage{amsmath}
\usepackage{amssymb}
\usepackage{graphicx}
\usepackage{hyperref}

\newcommand{\be}{\begin{equation}}
\newcommand{\ee}{\end{equation}}
\newcommand{\beq}{\begin{eqnarray}}
\newcommand{\eeq}{\end{eqnarray}}
\newcommand{\n}{\mathrm{n}}
\newcommand{\p}{\mathrm{p}}
\newcommand{\x}{\mathrm{x}}
\newcommand{\y}{\mathrm{y}}
\newcommand{\vv}{\mathrm{v}}
\newcommand{\R}{\mathcal{R}}
\newcommand{\B}{\mathcal{B}}

\newcommand{\en}{\varepsilon_\mathrm{n}}
\newcommand{\ep}{\varepsilon_\mathrm{p}}
\newcommand{\epsx}{\varepsilon_\mathrm{x}}

\newcommand{\mux}{\tilde{\mu}_\mathrm{x}}

\newcommand{\vb}{\mathbf{v}}
\newcommand{\vbx}{\mathbf{v}_{\mathrm{x}}}

\newcommand{\vxd}{{v}_{\mathrm{x}}}

\newcommand{\pxu}{{p}^{\mathrm{x}}}

\title[Hydrodynamical instabilities in superfluid neutron stars]{Hydrodynamical instabilities in superfluid neutron stars with background flows between the components.}
\author[V.~Khomenko, M.~Antonelli \& B.~Haskell]{V.~Khomenko, M.~Antonelli \& B.~Haskell \\
Nicolaus Copernicus Astronomical Center, Polish Academy of Sciences, ul. Bartycka 18, 00-716 Warsaw, Poland}

\begin{document}

\pagerange{\pageref{firstpage}--\pageref{lastpage}} \pubyear{2019}

\maketitle

\label{firstpage}

\begin{abstract}

The interiors of mature neutron stars are expected to be superfluid. Superfluidity of matter on the microscopic scale can have a number of large scale, potentially observable consequences, as the superfluid component of the star can now flow relative to the `normal' component that is tracked by electromagnetic emission. The most spectacular of such phenomena are pulsar glitches, sudden spin-ups observed in many pulsars. A background flow of a normal-fluid component with respect to the superfluid is also known to lead to a number of instabilities in laboratory superfluids, possibly leading to turbulence and modifying the nature of the mutual friction coupling between the two fluids. In this paper we consider modes of oscillation in the crust and core of a superfluid neutron star, by conducting a plane-wave analysis. We explicitly account for a background flow between the two components (as would be expected in the presence of pinning) and the entrainment, and we consider both standard (Hall-Vinen) and isotropic (Gorter-Mellink) forms of the mutual friction. We find that for standard mutual friction there are families of unstable inertial and sound waves both in the case of a counter-flow along the superfluid vortex axis and for counterflow perpendicular to the vortex axis and find that entrainment leads to a quantitative difference between instabilities in the crust and core of the star. For isotropic mutual friction we find no unstable modes, and speculate that instabilities in a straight vortex array may be linked to glitching behaviour, which then ceases until the turbulence has decayed.

\end{abstract}


\section{Introduction}

Neutron stars (NSs) are unique objects composed of matter at extremes of temperature, density and pressure that cannot be reproduced terrestrially. Describing their complex inner structure requires theoretical methods consistent with observations, which to date have been based on the analysis of the electromagnetic radiation emitted by the star, which carries mostly information on the processes in the magnetosphere or outermost, low density, layers. Despite the recent detection of gravitational waves emitted during a NS merger \citep{GW170817}, which is likely to deliver interesting constraints on the equation of state of matter at high densities in the near future \citep{LIGOEoS1}, direct data from the deep layers remains essentially inaccessible.

Nevertheless the state of matter in the stellar interior has a strong impact on the dynamics and observed astrophysical signals. In particular at high densities in the inner crust of a NS, neutrons can pair and form a superfluid, with protons also expected to be superconducting in the core (see \citet{HasSed} for a recent review). Superfluidity and superconductivity profoundly alter the dynamics of the components that can now flow relative to each other. 
The most striking phenomena connected to superfluidity are glitches, sudden spin-ups of the star,  that are instantaneous to the accuracy of the data, and are generally assumed to be due to the sudden re-coupling of the interior superfluid neutrons and the normal crust, which is tracked by the electromagnetic emission \citep{ai75, HaskellRev}. Glitches allow for an indirect probe of the NS interior, and can be used to obtain constraints on physical parameters of the star \citep{Alpar84b, Ho2015, Masse, Akbal17, Haskell2018a}. The superfluid component, however, also has a strong impact on the spectrum of oscillation of the NS, as the increase of degrees of freedom leads to an increase in the number of modes. This in turn will have observational consequences, as modes of oscillation can lead to gravitational wave emission \citep{Oscillo} and may be at the heart of the observed quasi-periodic oscillations in the tail of magnetar giant flares \citep{magnetarRev, Passamonti14, gabler16}.

The hydrodynamics of coupled superfluid and normal fluid systems plays a key role in the analysis of not only NSs, but also laboratory superfluids. In particular it is well known that thermal counterflow of the normal fluid with respect to the superfluid can generate turbulence, and that the vortex array will be disrupted by the so called Donnelly-Glaberson (DG) instability \citep{glab}. Many experimental and theoretical studies have recently investigated counterflow turbulence, and the link between the motion of the normal fluid and that of the superfluid \citep{Guo, Kivotides11, Guo17, Kivotides18}.  The instability is a general feature of superfluids with vorticity, and is thus likely to operate in NS interiors \citep{ts,waves}, as phenomena such as precession and Ekman flow can lead to large scale motions along the rotation axis  \citep{Peralta2005}.  The DG instability is not stabilized by the magnetic field, and could play an important role in a NS.\citep{Link2018}.

Very few experiments (see e.g. \citealt{SDB83, Eltsev03}), however, have dealt with instabilities due to counterflow coupled to rotation, which is a setup reminisicent of the neutron star interior, in which the normal component can rotate at a different rate with respect to the neutron superfluid. A notable exception is represented by the experiment of \citep{T1980}, who considered rotating spherical containers of superfluid He II, that abruptly changed angular velocity, and studied the relaxation of the fluid (see \citet{vE11} for a recent analysis of the experiment).

To make progress in the neutron star problem we adopt a hydrodynamical approach similar to that used in the study of superfluid He II, the so called HVBK two fluid model \citep{HV56,BK}, and based on the multi-fluid neutron star hydrodynamics proposed by  \citet{Mult}. This allows us to study modes of oscillation of the coupled fluids and investigate not only the mode spectrum, but also which modes may be driven unstable by the counterflow of the superfluid and normal components. Such instabilities may signal the transition to a turbulent state, and may both trigger a glitch and affect the response of the fluid after the event. An example of such trigger is the the Kelvin-Helmholtz instability on the isotropic-anisotropic interface between the $^1S_0$ and $^3P_2$ neutron superfluids close to the crust-core transition \citep{MM}. Another group of instabilities, namely two stream instabilities, may also be related to glitches since the entrainment effect could provide a sufficiently strong coupling for the instability to be astrophysically plausible \citep{EQS, Prix03, Trigger09, mike}. 
Instabilities in pinned superfluids have been studied by \citet{LinkCrust,LinkCore}, who considered a regime in which vortices are pinned either to nuclei in the crust, or to superconducting fluxtubes in the core of the star, with only a small number of them `creeping' out. On a larger scale, the global flow in a NS may also be susceptible to various instabilities in spherical Couette flow \citep{PerMel, Couette2006, Couette2008}, and such instabilities may also be related to timing noise in radio pulsars \citep{ltiming}, and precession of the rotational axis of NSs \citep{Gprec, PrecessionG}. 

In the present work a local three dimensional plane wave analysis of equations of motion for a multiconstituent fluid  \citep{Mult} is performed. We consider two fluids, the superfluid neutrons and a normal charge neutral fluid that consist of electrons and protons \citep{Mendell1991}. Dissipative coupling between the fluids is given by mutual friction, while a non-dissipative coupling is due the entrainment effect, that accounts for the reduced mobility of neutrons especially in the crust \citep{Prix2004, Chamel2017}. We start from the analysis of \citet{waves}, who analysed the oscillations of such a system, assuming the fluids to be locked in the background configuration, and in our calculation also allow for a velocity difference between the constituents in the background, and explicitly account for entrainment. Physically, such a background velocity difference will be built up if vortices are pinned in the crust or core of the star, and is crucial for our analysis.

% Perturbing quantities allow to find dispersion relations, starting with the simplest approaches with some quantities neglected. To study the properties of the associated waves, the new terms are being persistently added into the perturbed equations of motion. In order to make the analysis tractable, the dissipative and non dissipative mutual friction contribution is separated. 

\section {Two fluid hydrodynamics}

Our starting point will be the multifluid formalism of \cite{Mult}. We consider the hydrodynamical equations of motion for two dynamical degrees of freedom, the superfluid neutrons and a charge neutral fluid consisting of protons and electrons locked together by electromagnetic interactions on time scales shorter than those of interest for our problem. The momentum equations take the form:
\beq
\left.
\begin{aligned}
 &\left(\frac{\partial }{\partial t}+v_{\x}^{j}\nabla _j\right)\tilde{p}_{i}^{\x}+
 \varepsilon_\x w_{j}^{\y\x}\nabla_iv_{\x}^{j}+
\\ 
 &+\nabla_i\left(\Phi_R+\tilde{\mu}_\x\right)+ 2\epsilon_{ijk}\Omega^j v_{\x}^{k}=f_{i}^{\x}\, . 
\label{Euler}
\end{aligned}
\right.
\eeq

Indices $\x$ and $\y$ label the constituents, and the inequality $\x\neq\y$ is always understood to be true. The  proton-electron fluid will be denoted as $\p$ (and often referred to as the `proton' fluid, as electrons ensure charge neutrality, but only carry a small fraction of the inertia of the fluid) and superfluid neutrons, labeled as $\n$.  Indices $i,j,k$ label the spacial coordinates. Summation over repeated indices is implied (excluding the summation over constituent $\x$ and $\y$ indices). The velocity of constituent $\x$ is $v_{i}^\x$ while $w_{i}^{\y\x}=v_i^\y-v_i^\x$ is the difference between velocities of components. The angular velocity of the star in the background configuration is given by the vector $\Omega^i$, and we have included the centrifugal term in the potential $\Phi_R$, that in spherical coordinates can be written as 
\be
\Phi_R=\Phi-\frac{1}{2}\Omega^2 r^2\sin\theta^2 \, ,
\ee
where $\Phi$ is the gravitational potential which is given by the Poisson equation:
\be
\nabla^2\Phi=4\pi G\sum_\x \rho_\x \, .
\ee
 $\tilde{\mu}_\x = \mu_\x/m_\x$ is the chemical potential per unit mass, and $\rho_\x$ is the density of the $\x$ constituent. In the following we make the approximation $m_\p=m_\n=m$. Finally the momentum per unit mass $\tilde{p}_{i}^{\x}$, which due to entrainment is not aligned with the velocity $v^i_\x$, is:
\beq
\tilde{p}_{i}^{\x}=v_{i}^{\x}+\varepsilon_\x w_{i}^{\y\x} \, , \label{Momentum}
\eeq 
where $\varepsilon_\x$ is the entrainment coefficient. Assuming that individual species are conserved the continuity equations are:
\beq
\frac{\partial \rho_\x}{\partial t}+\nabla_j(\rho_\x v_{\x}^{j})=0 \, .\label{Continuo}
\eeq
The force $f_{i}^{\x}$ on the right hand side of the equations \eqref{Euler} is the vortex mediated mutual friction \citep{HV56}, and it represents an average of individual interactions between vortices and the normal fluid on the sub-hydrodynamical scale. As such its form depends strongly on the properties of the vortex configuration within the fluid.
For straight vortices and laminar flow, vortex mediated mutual friction force takes the form \citep{Andersson2006}:
\beq
 f_{i}^{\x}=\frac{\rho_\n}{\rho_\x}n_\vv \mathcal{B}^{'}\epsilon_{ijk}\kappa^j w_{\x\y}^{k}+\frac{\rho_\n}{\rho_\x}n_\vv \mathcal{B}\epsilon _{ijk}\hat{\kappa}^j\epsilon^{klm}\kappa_l  w_{m}^{\x\y} \, , 
 \label{eq1}
\eeq
where hats represent unit vectors, and the vector $\kappa^i=\kappa\hat{\Omega}^i$ points along the vortex array, which is co-linear with the rotation axis, and $\kappa=h/2m_\n$ is the quantum of circulation.
The vortex density per unit area, $n_\vv$, can be linked to the average large scale vorticity $\omega^i$ of the superfluid as
\beq
\kappa^i n_\vv=\omega^i=\, \epsilon^{ijk} \, \nabla_j \, \tilde{p}^\n_k  +2 \Omega^i\, ,
\eeq
which for two fluids rotating rigidly around a common axis with angular velocities $\Omega_\p$ and $\Omega_\n$, reduces to 
\beq
\kappa_i n_\vv=2\Omega_{i}^{\n}+2\varepsilon_\n(\Omega_{i}^{\p}-\Omega_{i}^{\n})  \, . 
\label{Fayn}
\eeq
For vanishing entrainment, or if the two fluids co-rotate, the above expression reduces to the standard Feynman-Onsager relation for rotating superfluids, $\kappa^i n_\vv=2\Omega^i_\n$.

The mutual friction parameters $\mathcal{B}$ and $\mathcal{B}^{'}$ in (\ref{eq1}) can be expressed in terms of dimensionless drag parameter $\mathcal{R}$ as:
\beq
\mathcal{B}=\frac{\mathcal{R}}{1+\mathcal{R}^2},\;\mbox{and}\;\; \mathcal{B}^{'} = \frac{\mathcal{R}^2}{1+\mathcal{R}^2}  \, ,
\eeq
where $\R$ encodes the microphysics of the dissipation processes that give rise to mutual friction in the stellar interior. %It equals to $R=\eta/\rho_s k$ and is connected with a drag force via the $\eta$. 

Theoretical calculations provide values for the drag parameter $\R \approx 10^{-4}$ for electron scattering on the vortex cores in the NS core \citep{Alpar1984} and $\R \approx 10^{-10}$ for phonon scattering in the crust \citep{Jones1990}. Higher value $\R \approx 1$ are expected due to Kelvons in the crust \citep{Jones1992, Epstein1992}, if vortices are moving rapidly past pinning sites. Using the two fluid model of \citet{Khomenko}, that allows for vortex accumulation and differential rotation, \citet{Haskell2018a} recently derived values of $\mathcal{R}$ from observations of glitches in Vela and Crab pulsar. They lies in the ranges $\B\approx 10^{-4}-10^{-3}$ for the Vela core and $\B\approx 10^{-5}-10^{-4}$ for the Crab crust. Similar results were obtained by \citep{Graber2018}, who also calculated the density dependence of the Kelvon mutual friction parameter.

In the presence of turbulence, however, the vortex array is disrupted, and a vortex tangle is likely to develop. In this case the form of the mutual friction in (\ref{eq1}) will no longer be appropriate. The form of the mutual friction to be used in the case of a polarized turbulent tangle in a neutron star is highly uncertain \citep{Andersson2007}, however in analogy to the cause of homogeneous isotropic turbulence in He-II, we will consider the form proposed by  \citet{GM}
\beq
f^{i}_\x 
\, = \, 
- \dfrac{\rho_\n}{ \rho_\x}  A_{GM} w_{\x\y}^2 \,  w_{\x\y}^i
 \, ,
\label{GM}
\eeq
where the parameter $A_{GM}$ has the dimensions of the inverse of a circulation and 
$w_{\x\y}^2 =w_{\x\y}^i w^{\x\y}_i$. As discussed by \cite{Andersson2007}, this form is essentially phenomenological and its relevance for neutron stars is still unclear. However, it is interesting to use this form of mutual friction as a completely different, physically motivated, alternative to \eqref{eq1}, to investigate the effects of fully developed turbulence on the mode structure of the star.

\section {Perturbed equations of motion}
\label{generalpert}
We begin our analysis by linearising the equations of motion in a frame rotating with angular velocity $\Omega=\Omega_\p$. Formally we expand a generic quantity $Q$ in a neighbourhood of a certain position $\mathbf{r}$, as
\beq
Q(\mathbf{r}, t) \, \approx \, Q_B(\mathbf{r}) + \delta Q (\mathbf{r},t) \, ,
\label{QB}
\eeq
where $Q_B$ is the background (time independent) value of the field, and the perturbation is assumed to be such that $|\delta Q|\ll |Q_B|$. With this in mind, the perturbed continuity equation follows from (\ref{Continuo}) and is:
\beq
\frac{\partial\delta \rho_\x }{\partial t}+\nabla ^j ( v_{j}^{\x}\delta\rho_\x+\rho_\x\delta  v_{j}^{\x})=0  \, ,
\label{PertCont}
\eeq
while from (\ref{Euler}) we obtain:
\be
\left.
\begin{aligned}
&( \partial_t +v_{\x}^{j}\nabla_j)(\delta v_{i}^{\x}(1-\varepsilon_\x)+w_{i}^{\y\x}\delta\varepsilon_\x +\varepsilon_\x \delta v_{i}^{\y})+\delta v_{\x}^{j}\nabla_j \tilde{p}_{i}^{\x}+
\\&\delta(\varepsilon_\x w_{j}^{\y\x} \nabla_i v_{\x}^{j})+
\nabla_i(\delta\tilde{\mu}_\x+\delta\Phi_R)+2\epsilon_{ijk}\Omega^j\delta v_{\x}^{k}=\delta f_{i}^{\x}  \, ,
\end{aligned}\right.\label{Eulerpert1}
\ee
where $\delta f_{i}^{\x}$ is a perturbed mutual friction contribution. In order to keep our equations tractable we apply the Cowling approximation, taking $\delta\Phi_R=0$. Using equation (\ref{Momentum}), we can thus expand (\ref{Eulerpert1}) to write
\be
\left.
\begin{aligned}
&( \partial_t +v_{\x}^{j}\nabla_j)
(\delta v_{i}^{\x}(1-\varepsilon_\x)+\varepsilon_\x \delta v_{i}^{\y}+w_{i}^{\y\x}\delta\varepsilon_\x)+\delta v_{\x}^{j}\nabla_j v_{i}^{\x}+\\
&\varepsilon_\x\delta v_{\x}^{j}\nabla_j w_{i}^{\y\x}+\varepsilon_\x  (\delta w_{j}^{\y\x} \nabla_i v_{\x}^{j}+w_{j}^{\y\x} \nabla_i \delta v_{\x}^{j})+\delta\varepsilon_\x w_j^{\y\x}\nabla_i v^j_\x+\\
&\nabla_i\delta\tilde{\mu}_\x+2\epsilon_{ijk}\Omega^j\delta v_{x}^{k}=\delta f_{i}^{x}  \, ,
\end{aligned}\right.\label {MAIN}
\ee
where
\be
\delta w^i_{\x\y}=\delta v^i_\x-\delta v^i_\y.
\ee

\subsection{Thermodynamical relations}
\label{termo}
\begin{figure}
  \centering
  \includegraphics[width=0.99\linewidth]{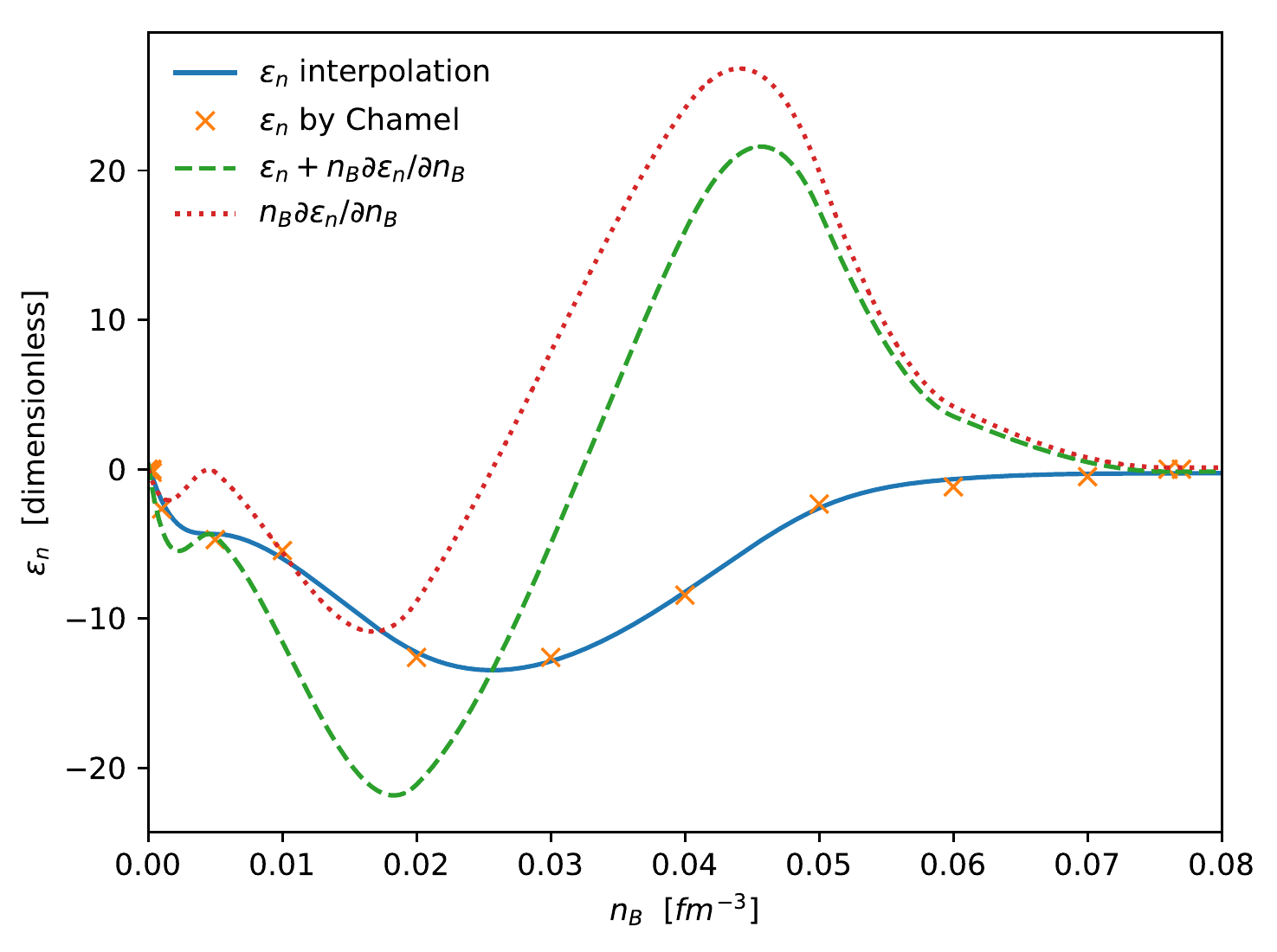}
  \caption{
Estimate of the quantity $\alpha_x$ as a function of the total baryon number density $n_B$ in the inner crust of a neutron star. We interpolate the orange crosses, that indicate the entrainment parameters $\epsx$ in the inner crust calculated by \citet{Chamel12}. The green dashed line is a rough estimate of $\alpha_\x$, the red dotted line quantifies the absolute error that arises by considering $\alpha_\x \approx \epsx$, according to equation \eqref{ennb}.}
\label{fig:entr}
\end{figure}

As it is clear from the form of the perturbation equations in the previous section, to compute the effect of a velocity perturbation we need to calculate perturbations of the chemical potential and of the entrainment. For a charge neutral two-fluid system it is known that a complete thermodynamic description in the zero temperature limit can be achieved by considering the two densities $\rho_\x$, the squared velocity lag between the two species $w_{\y\x}^2$ and the related intensive variables, namely the two chemical potentials per unit mass $\tilde{\mu}_\x$ and a dimensionless parameter $\alpha$ which regulates the strength of the conservative entrainment coupling between the two species \citep{Prix2004}.
In particular, the first law of thermodynamics can be expressed as \citep{andersson_comer2000,Andersson2006}
\beq
d E
\, = \, 
\tilde{\mu}_\x \, d\rho_\x + \tilde{\mu}_\y \, d\rho_\y + \alpha \, dw_{\y\x}^2 \, ,
\label{first_thermo}
\eeq 
where ${E}$ is the  energy density of the fluid mixture, while the intensive variables are given by
\beq
\tilde{\mu}_\x \, = \, \frac{\partial {E}}{\partial \rho_\x}\Big{\vert}_{\rho_\y w_{\y\x}^{2}} \, 
\quad \text{and} \quad
\alpha \, = \, \frac{\partial {E}}{\partial w_{\y\x}^{2}}{\Big\vert}_{\rho_\x\rho_\y} \, .
\label{first_thermo_def}
\eeq 
The parameter $\alpha$ has the dimensions of a mass density and is related to the usual dimensionless entrainment parameters $\varepsilon_\x$ via \citep{Prix2004,CarterChamel2006}
\beq
\varepsilon_\x = \frac{2\alpha}{\rho_\x} \, .
\label{chamel06}
\eeq
Therefore, the variation $\delta\epsx$ is related to the variations of the more basic thermodynamic quantities $\alpha$ and $\tilde{\mu}_\x$ as
\beq
\delta \epsx 
\, = \,  
\frac{2 \, \delta\alpha}{\rho_x} - \frac{\epsx \, \delta \rho_x}{\rho_x}  \, .
\label{deps}
\eeq
While for the chemical potential we need to calculate
\beq
\delta \tilde{\mu}_{\x}=\frac{\partial \tilde{\mu}_{\x}}{\partial \rho_\x}\delta \rho_\x+\frac{\partial \tilde{\mu}_{\x}}{\partial \rho_\y}\delta \rho_\y+\frac{\partial \tilde{\mu}_{\x}}{\partial w_{\y\x}^{2}}\delta (w_{\y\x}^{2}) \,.\label {eq4}
\eeq
The derivatives of chemical potential with respect to the squared background velocity may be computed using the first law of thermodynamics in \eqref{first_thermo}, which leads to:
\be
\frac{\partial \tilde{\mu}_\x}{\partial w_{\y\x}^2}=\frac{\partial^2 E}{\partial w_{\y\x}^2 \partial \rho_\x}=\frac{\partial }{\partial \rho_\x}\left(\frac{\partial E}{\partial w_{\y\x}^2}\right)=\frac{\partial \alpha}{\partial \rho_\x}\, ,
\ee
which, from (\ref{chamel06}) gives:
\be
\frac{\partial \tilde{\mu}_\x}{\partial w_{\y\x}^2}=\frac{1}{2}\varepsilon_\x+ \frac{1}{2}\rho_\x \frac{\partial \varepsilon_\x}{\partial \rho_\x}\, .\label {rv1}
\ee
We are now ready to calculate perturbations of the entrainment and chemical potential. To do this we introduce the sound velocities $c_\x$, the chemical couplings $C_\x$ \citep{Andersson2001} and two additional parameters related to the dependence on the lag of $\alpha$ and the chemical potentials,
\begin{align}
\begin{split}
c_\x^2 \, &=\, \rho_\x \, \dfrac{\partial  \tilde{\mu}_\x }{\partial \rho_\x }
\\
C_\x \, &=\, \rho_\y \dfrac{\partial  \tilde{\mu}_\x}{\partial \rho_\y }
 \, =\,  \rho_\y \dfrac{\partial  \tilde{\mu}_\y}{\partial \rho_\x }
\\
A_\x \, &=\, \dfrac{4}{\rho_\x}\dfrac{\partial \alpha}{\partial  w^2_{\n\p}}
\\
\alpha_\x \, &=\, 2\dfrac{\partial \alpha}{\partial \rho_\x} 
\,=\, 2\dfrac{\partial \tilde{\mu}_\x}{\partial w^2_{np}}
\,=\, \epsx +\rho_\x \dfrac{\partial  \epsx }{\partial \rho_\x } \, .
\end{split}
\label{definitions}
\end{align}
Note that our definition of $c_\x^2$  coincides with the one used by \cite{waves}. In our more general analysis, we let the entrainment vary according to the thermodynamic relations, which forces us to introduce also $A_\x$ and $\alpha_\x$, at least at this formal level. There are no estimates in the literature for the parameters $A_\x$ in a neutron star, so in the following we will neglect it and set $A_\x=0$ in practical examples. The parameter $C_\x$ is also highly uncertain and generally computed from phenomenological models \citep{EQS}, while for $\alpha_\x$ one can obtain an estimate in the inner crust, shown in Fig \ref{fig:entr}, by assuming that the ratio between the densities of the two species is frozen at its chemical equilibrium value. In this case one has that
\beq
\epsx +\rho_\x \dfrac{\partial  \epsx }{\partial \rho_\x} 
\, \approx \,  
\epsx +n_B \dfrac{\partial  \epsx }{\partial n_B} 
\label{ennb}
\eeq
where $n_B$ represents the total baryon number density. This rough argument gives a plausibility interval for the unknown values of $\alpha$.

%However, given the large uncertainties in the core, and for computational ease, we will, however, take $\alpha_\x=\varepsilon_\x$ in the following, unless explicitly stated otherwise.

Thanks to the fact that $\delta w_{\x\y}^2 = 2 (v^\x_i-v^\y_i) (\delta v_\x^i - \delta v_\y^i)$, 
the variation of the chemical potentials can be conveniently written as

\begin{align}
\begin{split}
\delta \tilde{\mu}_\x 
&  \, = \, 
\dfrac{\partial \tilde{\mu}_\x }{\partial \rho_\x} \delta \rho_\x
+
\dfrac{\partial \tilde{\mu}_\x }{\partial \rho_\y} \delta \rho_\y
+
\dfrac{\partial \tilde{\mu}_\x }{\partial w_{\y\x}^2} \delta w_{\y\x}^2 \,  
\\
  & \, = \, 
c_\x^2 \, \dfrac{\delta \rho_\x}{\rho_\x}
+
C_\x \, \dfrac{\delta \rho_\y}{\rho_\y}
+
\alpha_\x \, w^{\y\x}_i \, \delta w_{\y\x}^i \, .
\end{split}
\label{mutilde}
\end{align}

Following the same procedure, we can write for the entrainment:
\begin{align}
\begin{split}
\delta \alpha
& \, = \, 
\dfrac{\partial \alpha}{\partial \rho_\x} \delta \rho_\x
+
\dfrac{\partial \alpha}{\partial \rho_\y} \delta \rho_\y
+
\dfrac{\partial \alpha}{\partial w_{\y\x}^2} \delta w_{\y\x}^2
\\
  & \, = \, 
\dfrac{\alpha_\x \rho_\x}{2}   \dfrac{\delta \rho_\x}{\rho_\x}
+
\dfrac{\alpha_\y \rho_\y}{2}   \dfrac{\delta \rho_\y}{\rho_\y}
+
\dfrac{A_\x \rho_\x}{2} \, w^{\y\x}_i \, \delta w_{\y\x}^i \, .
\end{split}
\label{dalpha}
\end{align}
which combined with equations \eqref{mutilde} and \eqref{dalpha} with \eqref{deps}, leads to
\beq
\delta \epsx 
\, = \, 
A_\x w_{\y\x}^i \delta w^{\y\x}_i + (\alpha_\x -\varepsilon_\x ) \dfrac{\delta \rho_\x}{\rho_\x} + \alpha_\y\dfrac{\delta \rho_\y}{\rho_\x} \, .
\label{deltaepsilon}
\eeq

\subsection{Perturbing the mutual friction}

The next step is to perturb the mutual friction. Let us start by considering the anisotropic form valid for a straight vortex array, given in (\ref{eq1}). In this case it is 
\beq
\left.
\begin{aligned}
&\delta f_{i}^{\x}\!=\!\delta \!\!\left(\frac{\rho_\n}{\rho_\x}\right)\!\![n_\vv \kappa^j \mathcal{B}^{'}\epsilon _{ijk}w_{\x\y}^{k}+n_\vv \kappa_l \mathcal{B}\epsilon _{ijk}\hat{\kappa}^j
\epsilon^{klm}w_{m}^{\x\y}]+\\
&\frac{\rho_\n}{\rho_\x}n_\vv \kappa^j \mathcal{B}^{'}\varepsilon _{ijk}\delta w_{\x\y}^{k}+\frac{\rho_\n}{\rho_\x}n_\vv \kappa_l \mathcal{B}\varepsilon _{ijk}\hat{\kappa}^j\varepsilon^{klm} \delta w_{m}^{\x\y}+\\
& \frac{\rho_\n}{\rho_\x} \delta(n_\vv \kappa^j) \mathcal{B}^{'}\varepsilon _{ijk} w_{\x\y}^{k}+\frac{\rho_\n}{\rho_\x}\delta (n_\vv \kappa_l) \mathcal{B}\varepsilon _{ijk}\hat{k}^j\epsilon^{klm} w_{m}^{\x\y}+\\
& \frac{\rho_\n}{\rho_\x}n_\vv \kappa_l \mathcal{B}\varepsilon _{ijk} \delta (\hat{\kappa}^j) \varepsilon^{klm} w_{m}^{\x\y}\, .
\end{aligned}\right.
\label {MFD}
\eeq
where we have assumed that $\delta\mathcal{B}=\delta\mathcal{B}^{'}=0$. In principle one could vary also these quantities, but given the large uncertainties on these coefficients, and to simplify our analysis, we choose to ignore their perturbations. In order to discuss the terms in a more tractable form, and keep track of their physical origin, we separate the expression above in the following way:
\beq
\delta f_{i}^{\x} = (\delta f_{i})_{sound}+(\delta f_{i})_{w}+(\delta f_{i})_{n_\vv \kappa}+(\delta f_{i})_{\hat{\kappa}}\, .
\eeq
The first term, $(\delta f_{i})_{sound}$ connects sound waves and the mutual friction and takes the form
\begin{multline}
(\delta f_{i})_{sound}
=
\left(\frac{\delta \rho_\n}{\rho_\x} - \frac{\rho_\n}{\rho_{\x}^2}\delta\rho_\x\right)(n_\vv \kappa^j \mathcal{B}^{'}\epsilon _{ijk}w_{\x\y}^{k}+  
\\
+n_\vv \kappa_l \mathcal{B}\epsilon _{ijk}\hat{\kappa}^j
\varepsilon^{klm}w_{m}^{\x\y})\, .
\end{multline}
The term, containing the perturbations of the background velocity, which would exist also if the fluids co-rotate in the background, is:
\be
(\delta f_{i})_{w}=\frac{\rho_\n}{\rho_\x}n_\vv (\kappa^j \mathcal{B}^{'}\epsilon _{ijk}\delta w_{\x\y}^{k}+ \kappa_l \mathcal{B}\epsilon _{ijk}\hat{\kappa}^j\epsilon^{klm} \delta w_{m}^{\x\y}) \,.
\ee
The final two terms, $(\delta f_{i})_{n_\vv \kappa}$ and $(\delta f_{i})_{\hat{\kappa}}$ depend on variations of the vorticity, both in magnitude and direction, and to calculate them we need to perturb the vorticity $\omega^i=\kappa^i n_\vv$, so that
\be
\delta(\kappa^i n_\vv)=\delta\omega^i=\epsilon^{ijk} \nabla_j \delta \tilde{p}^\n_k\,\,
\ee
which using 
\be
\delta \tilde{p}^i_\x=(1-\varepsilon_\x)\delta v_\x^i+\varepsilon_\x \delta v^i_\y+w^i_{\y\x}\delta\varepsilon_\x\, ,
\label{deltap1}
\ee
gives
\be
\left.
\begin{aligned}
\delta\omega^i=&(1-\varepsilon_\n)\epsilon^{ijk}\nabla_j\delta v_k^\n +\varepsilon_\n\epsilon^{ijk}\nabla_j\delta v_k^\p\\
&+\epsilon^{ijk}\nabla_j(\delta\varepsilon_\n) w_k^{\p\n}\,.
\end{aligned}\right.
\label{dw_gen}
\ee
We also need to perturb the vorticity unit vector $\hat{\kappa}^i=\hat{\omega}^i$,  which can be done readily in terms of a projection operator $\perp$ that projects orthogonally to the vortex lines:
\beq
\delta \hat{\kappa}^i\,=\,\delta \hat{ \omega}^i
\, = \, 
\frac{1}{|\boldsymbol{\omega}|}
\perp^i_a \, \delta \omega^a 
\, = \, 
\frac{1}{|\boldsymbol{\omega}|}
(\delta^i_a - \hat{ \omega}^i \hat{ \omega}_a )  \, \delta \omega^a \, .
\label{perp}
\eeq
We can thus write
\be
\left.
\begin{aligned}
(\delta f_{i})_{n_\vv \kappa}=&-\frac{\rho_\n}{\rho_\x} \mathcal{B}^{'} (w_{\x\y}^{k}\nabla_k {\delta \tilde{p}_{\n}^{i}}-w_{\x\y}^{k}\nabla_i {\delta \tilde{p}_{k}^{\n}})+\\
&\frac{\rho_\n}{\rho_\x} \mathcal{B} (\hat{\kappa}^m w_{m}^{\x\y} \epsilon_{ith} \nabla^t \delta\tilde{p}_{\n}^{h} -w_{i}^{\x\y}\hat{\kappa}^l\epsilon_{lth} \nabla^t \delta\tilde{p}_{\n}^{h})\, ,
\end{aligned}\right.
\ee
and
\beq
(\delta f_{i})_{\hat{\kappa}}=\frac{\rho_\n}{\rho_\x} \mathcal{B} \hat{\kappa} _{i}(w_{\x\y}^s \epsilon_{slm} \nabla^l {\delta \tilde{p}_{\n}^{m}}- w^{\x\y}_{q} \hat{\kappa}^q \epsilon_{tlm} \hat{\kappa}^t \nabla^l {\delta \tilde{p}_{\n}^{m}}) \, ,
\eeq
which, together with (\ref{deltap1}) completes our calculation of the perturbed mutual friction. Before moving on, let us remark that we could have obtained our expression directly in terms of the perturbed vorticity $\delta\omega^i$, by noting that the term $\epsilon_{ijk} \hat{\omega}^j \epsilon^{klm} \omega_l  {w}_m^{\y\x}$ in the expression in (\ref{eq1}) for the mutual friction, can be written in terms of the projection operator in (\ref{perp}) as
\beq
\epsilon_{ijk} \hat{{\omega}}^j \epsilon^{klm} \omega_l  {w}_m^{\y\x}
\, = \, 
-|\boldsymbol{\omega}|     \perp_i^j {w}_j^{\y\x}   \, .
\eeq
and that
\begin{multline}
-\delta \,  (\, |\boldsymbol{\omega}|     \perp_i^j {w}_j^{\y\x} \, )
\, = \, 
\delta{\omega}_i \, (\hat{{\omega}}^j w_j^{\y\x})
+
\hat{{\omega}}_i  \, (  w^j_{\y\x} \delta{{\omega}}_j) +
\\
-(\hat{{\omega}}^j  \delta{\omega}_j) (\perp_i^k w_k^{\y\x})   
- |\boldsymbol{\omega}| (\perp_i^k \delta{w}_k^{\y\x})   
\, .
\label{cavolino}
\end{multline}

\subsubsection{Isotropic mutual friction}

We consider also perturbations of the isotropic mutual friction force in (\ref{GM}), proposed for the case in which we have fully developed turbulence and an isotropic vortex tangle. In this case the perturbations of the vorticity play no role and from equation (\ref{GM})  we have:
\be
\left.
\begin{aligned}
\delta {f}_i^{\x (GM)} \,=& \, - \frac{\rho_\n}{\rho_\x} A_{GM} \left\{|\mathbf{w}_{\x\y}|^2 \, \delta{w}_i^{\x\y} + 2 ( w_{\x\y}^j\delta w_j^{\x\y})  \,{w_i}^{\x\y}\right\} -\\
&\delta\left(\frac{\rho_\n}{\rho_\x}\right)A_{GM} |\mathbf{w}_{\x\y}|^2 \,  w_{\x\y}^i\, .
\end{aligned}
\right. \label{pertGM}
\ee

\section{Plane wave analysis}
\label{planew}

We are now ready to study the oscillation spectrum of our neutron star model, and to do so we will make the plane wave approximation. To do this we assume that our oscillations can be described as plane waves in a neighbourhood $\mathbf{r} + \mathbf{x}$ of the point $\mathbf{r}$ at which we calculated our background quantities  $Q_B(\mathbf{r})$, so that 
\beq
Q(\mathbf{r} + \mathbf{x}, t) \, \approx \, Q_B(\mathbf{r}) + \delta Q_{\mathbf{r}} (\mathbf{x},t) \, ,
\label{QB}
\eeq
with
\beq
\delta Q_\mathbf{r} \, = \, \bar{Q} \, e^{i\left( {k}_j{x}^j - \omega t \right)} \, .
\label{Qplane}
\eeq

%%%%%%%%%%%%%%%%
\begin{figure}
  \centering
  \includegraphics[width=0.65\linewidth]{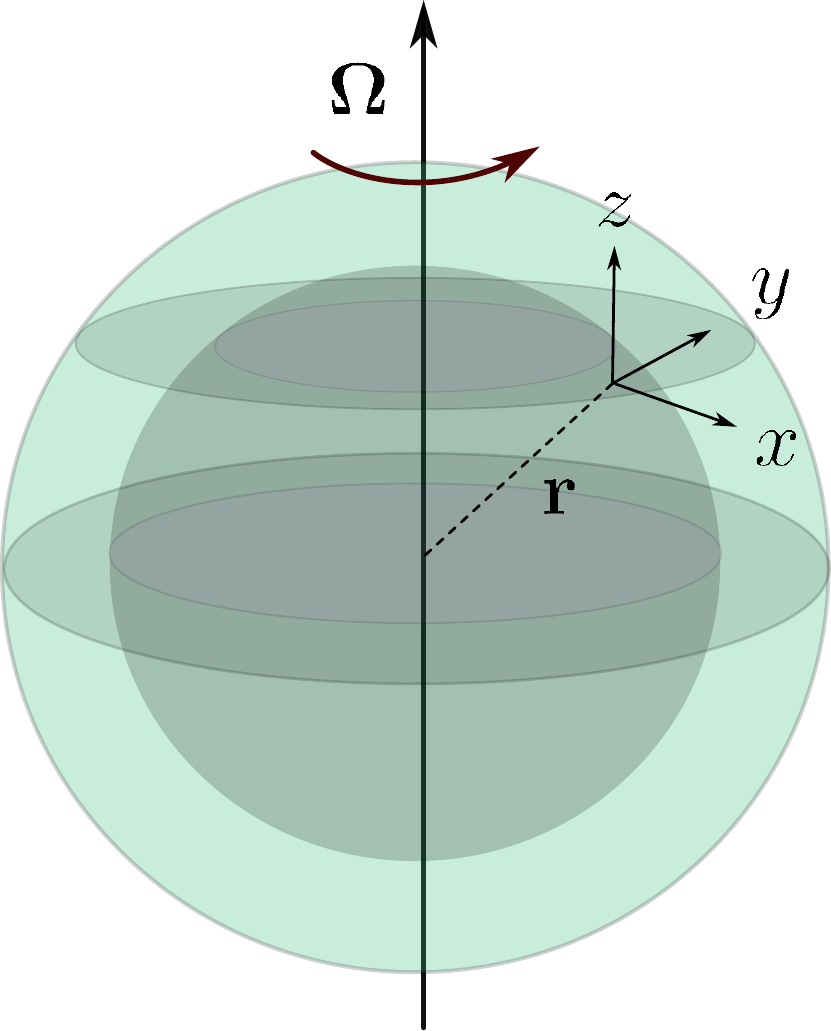}
  \caption{
  Sketch of the geometrical definitions used: the vector $\mathbf r$ indicates the position of a small sample of matter inside the star. The local Cartesian system of coordinates is chosen to be right-handed in such a way that the $z$-axis is parallel to the direction identified by $\mathbf \Omega$, the $x$-axis defines the direction of the cylindrical radius and the $y$-axis is locally parallel to the azimuthal direction. The coordinates $x$, $y$ and $z$ identify the position $\mathbf{x}$ introduced in equation \eqref{QB}.
  }
 \label{fig:geo}
\end{figure}
%%%%%%%%%%%%%%%%%%%

We assume that $|\mathbf{x}| \ll |\mathbf{r}|$, so that all gradients of background quantities are ignored in our perturbation equations. We use a local cartesian coordinate system, as sketched in figure \ref{fig:geo}, which is taken to be co-rotating with the normal component, such that $\Omega=\Omega_\p$ and $v_\p=0$ in the following. This naturally implies that 
\be
w^i_{\n\p}=v^i_\n \, .
\ee
Each perturbed quantity $\delta Q$ can be expressed in terms of the two fields  $\delta v^i_\n$ and $\delta v^i_\p$ by means of the continuity equations and of the thermodynamic relations and it will be convenient to  express the amplitude $\bar{Q}$ as:
\beq
\bar{Q} \, = \,   Q^\n_a \, \bar{v}_\n^a  \, + \,   Q^\p_a  \, \bar{v}_\p^a  \, .
\label{Qgeneral}
\eeq
Following this procedure, the perturbed version of the Euler-like equations (\ref{Euler}) can be used to define a linear system of six equations for the six independent amplitudes $\bar{v}_\x^i$ which has the form
\beq
\sum_{\y^{'}=\,\n,\p} \, M_{\,i\,j}^{\,\x \,\y^{'}} \bar{v}^{\,j}_{\y^{'}}\, .
\label{general}
\eeq
%\beq
%\sum_{(\x' \,  j)} \, M_{(\x \, i)(\x' \,  j)} \, \bar{v}_{\x'}^j \, = \, 0 \, ,
%\label{general}
%\eeq
%
%where the multi-index $(\x i)$ runs over the six combinations 
%$\{ n\,x, \,n\,y, \,n\,z, \,p\,x, \,p\,y, \,p\,z\}$. 
The only way to have non trivial solutions of the above equation is to impose that  $\det{(M)}=0$. The determinant of (\ref{general}) is in general a high degree polynomial in the components of the wave vector $\mathbf{k}$ and of the frequency $\omega$ and requires a numerical analysis. Particular cases that can be studied analytically will, however, be presented in the following sections. 

To calculate the components of the matrix in (\ref{general}) first of all we insert the plane wave ansatz in the continuity equation (\ref{PertCont}), to obtain
\beq
\bar{\rho}_\x
\, = \, {\rho}_i^{\x\x}\, \bar{v}^i_\x \, ,
\quad \text{where} \quad
{\rho}_i^{\x\x} \, = \,\dfrac{ \rho_\x \, k_i }{ \omega - k_i v^i_\x }
 \, .
\label{rho_amplitude}
\eeq
We also need the plane wave perturbations of the chemical potential and entrainment, which from (\ref{mutilde}) and (\ref{deltaepsilon}), are:
%\beq
%\bar{\mu}_\x  \, = \, \mu^{\x \n}_a \, \bar{v}_\n^a  \, + \,    \mu^{\x \p}_a  \, \bar{v}_\p^a  
%\label{muu}
%\eeq
%
%and
%
%\beq
%\bar{\varepsilon}_\x \, = \, \varepsilon^{\x \n}_a \, \bar{v}_\n^a  \, 
%+ \, \varepsilon^{\x \p}_a  \, \bar{v}_\p^a  \, \, .
%\label{espss}
%\eeq
%Substituting the plane wave ansatz in the expression in (\ref{mutilde}), we obtain 
\beq
\bar{\mu}_\x 
 \, = \, 
\left[  c_\x^2  \,  \dfrac{\rho^{\x\x}_i}{\rho_\x}+ \alpha_\x v_{\x\y}^i \right] \bar{v}_\x^i
+
\left[ C_\x \,  \dfrac{\rho^{\y\y}_i}{\rho_\y} - \alpha_\x v_{\x\y}^i \right] \bar{v}^\y_i 
\, ,
\label{mubar}
\eeq
and
\beq
\bar{\varepsilon}_\x 
= 
\left[ (\alpha_\x-\epsx) \dfrac{  {\rho}_i^{\x\x} }{{\rho}_{\x}}+A_\x {v}_{\x\y}^i   \right]  \bar{v}^{\x}_i
+
\left[\alpha_\y  \dfrac{\rho_i^{\y\y}}{\rho_\x}-A_\x {v}_{\x\y}^i  \right]  \bar{v}^{\y}_i \, .
\label{amp_entr}
\eeq
%
%that, using (\ref{rho_amplitude}), allows us to conclude that
%
%\beq
%\begin{split}
%\mu^{\x\x}_i
%&\, = \, 
%  \dfrac{c_\x^2\, k_i}{\omega - {k}_j v^j_\x} + \alpha_\x v^{\x\y}_i
%\\
%\mu^{\x\y}_i
%&\, = \, 
%\dfrac{C_\x \, k_i}{\omega - {k}_j v^j_\y} - \alpha_\x v^{\x\y}_i
 %\, .
%\label{amp_mu}
%\end{split}
%\eeq
%Similarly, from (\ref{deltaepsilon}), we obtain

% 
%which, together with \eqref{rho_amplitude}, gives:
%
%\beq
%\begin{split}
%\varepsilon^{\x\x}_i
%&\, = \, 
% (\alpha_\x-\epsx) \dfrac{  \, k_i}{\omega - {k}_j v^j_\x}+A_\x {v}^{\x\y}_ii   
%\\
%\varepsilon^{\x\y}_i
%&\, = \, 
%\dfrac{\alpha_\y \rho_\y}{\rho_x} \dfrac{ k_i}{\omega - {k}_j v^j_\y }-A_\x {v}^{\x\y}_i   \, .
%\label{amp_entr2}
%\end{split}
%\eeq
With the above expressions at hand we can now compute the plane wave perturbations of the Euler equations in (\ref{MAIN}). The full result is complex, but a procedure to obtain all the variations in the presence of a background lag and with no approximations is given in Appendix \ref{appendixequations}. 
It is, however, instructive to study the case in which we neglect perturbations of the entrainment and set $\delta\varepsilon_\x=0$ and also neglect terms involving $\partial\varepsilon_\x/\partial\rho_\x$, so that one has $\alpha_\x\approx \varepsilon_\x$ in \eqref{mubar}. In this case the Euler equations can be written as
\be
\left.
\begin{aligned}
&i(k_j v_{\x}^j-\omega)[\bar{v}_{i}^{\x}(1-\varepsilon_\x)+\varepsilon_\x\bar {v_{i}^{\y}}]+2\epsilon_{ijk}\Omega^j\bar{v_{x}^{k}}-\\
&i k_i(c_\x^2 \frac{k_j \bar{v}_{x}^{j}}{k_j v_{\x}^{j}-i\omega}+C_\y\frac{k_j \bar{v}_{y}^{j}}{k_j v_{\y}^{j}-i\omega}-\varepsilon_\x\bar{v}_{y}^{j}w^{\y\x}_{j})=\delta f_i \, ,
\label{eqLHS}
\end{aligned}
\right.
\ee
where defining $\bar{p}_{\n}^{k} = \bar{v}_{\n}^k+\varepsilon_\n \bar{w}_{\p\n}^{k}$, we have
\be
\left.
\begin{aligned}
\delta f_i=& \left(\frac{ \bar{\rho}_\n}{\rho_\x} - \frac{\rho_\n}{\rho_{\x}^2}\bar{\rho}_\x\right)n_\vv \kappa^j \mathcal{B}^{'}\epsilon _{ijk}w_{\x\y}^{k}+\\
&\left(\frac{ \bar{\rho}_\n}{\rho_\x} - \frac{\rho_\n}{\rho_{\x}^2}\bar{\rho}_\x\right) n_\vv \kappa_l \mathcal{B}\epsilon _{ijk}\hat{\kappa}^j
\epsilon^{klm}w_{m}^{\x\y}+\\
&\frac{\rho_\n}{\rho_\x}n_\vv \kappa^j \mathcal{B}^{'}\epsilon _{ijk}\bar{w}_{\x\y}^{k}+\frac{\rho_\n}{\rho_\x}n_\vv \kappa_l \mathcal{B}\epsilon _{ijk}\hat{\kappa}^j\epsilon^{klm} \bar{w}_{m}^{\x\y}-\\
&\frac{\rho_\n}{\rho_\x} \mathcal{B}^{'} (i w_{\x\y}^{k} k_k \bar{p}^{\n}_{i}-i w_{\x\y}^{k} k_i \bar{p}_{k}^{\n})+\\
&\frac{\rho_\n}{\rho_\x} \mathcal{B} (i\hat{\kappa}^m w_{m}^{\x\y} \epsilon_{ith} k^t \bar{p}_{n}^{h} -i w_{i}^{\x\y}\hat{\kappa}^l\epsilon_{lth}k^t \bar{p}_{\n}^{h})+\\
&\frac{\rho_\n}{\rho_\x} \mathcal{B} \delta _{i}^{z}(i w_{\x\y}^s \epsilon_{slm} k^l \bar {p}_{\n}^{m} - i w^{\x\y}_{z} \epsilon_{zlm} k^l \bar {p}_{\n}^{m}) \, .
\end{aligned}
\right.
\label{eqRHS}
\ee

\section{Waves in the absence of mutual friction}

In this section we consider some simple examples where progress can be made analytically, before moving on to the numerical results for the full set of equations. In particular we will first consider sound waves and inertial waves in the limit of vanishing mutual friction, and show how mutual friction can damp or drive some of the modes instable. The approximations will be outlined for each case as we proceed.

\subsection{Sound waves}

Let us begin by considering pure sound waves. If we set $\Omega=\varepsilon_\x=w_{\n\p}=v_\n=0$ in (\ref{eqLHS}) and (\ref{eqRHS}) the dispersion relation follows from 
\be
\omega^2(c_\p^2 k^2 - \omega^2)(c_{\n}^{2} k^2-\omega^2)=0
\ee
 we recover simply two branches of sound waves, one for each fluid
\beq
&&\omega=\pm kc_{\n}\\
&&\omega=\pm kc_{\p}
\eeq
Including a background flow, still with $\Omega=\varepsilon_\x=0$ does not couple the fluids, and simply introduces a correction:
\beq
&&\omega =  (v^\n_i k^i) \\
&&\omega=\pm kc_{\n} + (v^\n_i k^i)\\
&&\omega=\pm kc_{\p}
\eeq

\subsubsection{Chemical coupling and entrainment}

We now move on to considering the chemical coupling between the two fluids in the regime in which $\varepsilon_\x=w_{\n\p}=v_\n=0$ and $\Omega=0$. In this case the dispersion relation follows from

\begin{multline}
\omega^4  \, \xi \, - \omega^2  k^2( \, C_\varepsilon+c_\varepsilon^2 \, )
+k^4(c_\n^2 c_\p^2 - C_n C_p) = 0  \, ,
\label{no_rott}
\end{multline}
where we have defined
\beq
C_\varepsilon \, =\,  C_\n \ep +C_\p \ep \,  =\,  4 \, \alpha \,  \dfrac{\partial^2 E}{\partial \rho_n \partial \rho_\p} \,.
\eeq
and
\beq
c_\varepsilon^2 \, =\,  (1-\en)c_\p^2 + (1-\ep)c_\n^2 \, ,
\label{cc}
\eeq
as well as the dimensionless parameter
\beq
\xi \, =\,  1-\en-\ep \, .
\label{xie}
\eeq

The roots of (\ref{no_rott}) are organized in four branches that have the form 
\beq
\omega \, = \, \pm \, k \, c_{1}
\qquad \qquad
\omega \, = \, \pm \, k \, c_{2}
\, ,
\eeq
where the constants $c_{1}$ and $c_{2}$ are two velocities for entrainment-coupled sound waves, 
\beq
\begin{split}
& c^2_{1}  
\, = \, 
\dfrac{c^2_\varepsilon +C_\varepsilon + \Delta }{2 \xi}
\\
& c^2_{2}  
\, = \, 
\dfrac{c^2_\varepsilon +C_\varepsilon - \Delta }{2 \xi}\, ,
\end{split}
\label{soundd}
\eeq
with
\be
\Delta 
\, = \, 
\sqrt{ (c^2_\varepsilon +C_\varepsilon)^2 -4 \xi (c_\n^2 c_\p^2 - C_\n C_\p)}
\, .
\ee
It is immediate to see that, in the limit of no entrainment and no chemical coupling, $c_{1}$ and $c_{2}$ are exactly equal to $c_n$ and $c_p$ of pure one-component sound waves.
To grasp the physical effect of chemical coupling we furthermore require that $c_\x^2 \gg C_{\x},C_{\y} $. In this limit it is useful to define $\Delta_0$ as the limit of $\Delta$ when $C_{\x}=0$,
\beq
& \Delta_0 
\, = \, 
\sqrt{ c^2_\varepsilon -4 \xi c_\n^2 c_\p^2 } \, ,
\label{delta0}
\eeq
so that the the limit of weak chemical coupling can be equally seen as  $C_{x}/\Delta_0 \ll 1$.
In this case, the velocities in \eqref{sound} can be expanded  up to the second order in the chemical couplings as
\begin{multline}
c^2_{1,2}  
\, = \, 
\dfrac{c^2_\varepsilon +C_\varepsilon \pm \Delta_0 }{2 \xi}
\left( 1 \pm \dfrac{C_\varepsilon}{\Delta_0} \right) +
\\
\pm \dfrac{C_\n C_\p}{\Delta_0^2} \dfrac{ c_\varepsilon^4 - 4 \xi c^2_\n c^2_\p}{\Delta_0}
\mp \dfrac{C^2_\varepsilon}{\Delta_0^2} \dfrac{ c^2_\n c^2_\p}{\Delta_0} + O(C_\x^4)
\label{sound_C}
\end{multline}
In the limit of $\varepsilon_\n=\varepsilon_\p$, and using the relation $C_\p = \frac{\rho_\p}{\rho_\n}C_\n$ we recover the result \citep{waves}:

\beq
\omega = \pm k c_\n[1+\frac{\rho_\p}{2\rho_\n}\frac{C_\n^2}{c_\n^2(c_\n^2-c_\p^2)}] \\
\omega = \pm k c_\p[1+\frac{\rho_\p}{2\rho_\n}\frac{C_\n^2}{c_\p^2(c_\p^2-c_\n^2)}]
\eeq
	 However, retaining the entrainment terms, the lowest order correction to the dispersion relation comes the term that is linear in  $C_\varepsilon$: when entrainment is considered, the effect of the chemical coupling on sound waves is enhanced. Note that this result is independent of the assumption $\delta\varepsilon_\x=0$, and remains the same also in the more general case in which perturbations of the entrainment are considered.

\subsubsection{Including rotation}

We now consider the effect of rotation on the sound waves, in the limit of no chemical coupling, i.e. we take $C_\p=C_\n=w_{\n\p}=v_{\n}=0$ in the background. In this case the dispersion relation is still too complex to solve for a general case. However if we consider sound waves with a dispersion relation of the form  $\omega \sim c_\x k$, we expect that $\omega \gg \Omega$ even for the smallest possible $k$, which is of the order of the inverse of the stellar radius $R^{-1}$ (i.e. the sound velocity in nuclear matter is such that $c_\x \gg \Omega R$). Therefore, it is a very good approximation to expand the two branches of $\omega^2(k)$ to the lowest order in $\Omega$: 
\beq
\omega^2(k) \, = \, k^2 \, c^2_{1,2} + f_{1,2} \, \Omega^2  + O(\Omega^4)
\, ,
\eeq
where
\beq
c^2_{1,2}  
\, = \, 
\dfrac{c^2_\varepsilon \pm 
	\Delta_0}{2 \xi} 
\qquad \quad
\Delta_0 = \sqrt{ c^4_\varepsilon -4 \xi c_\n^2 c_\p^2 }
\, ,
\label{sound}
\eeq
where $c_1$ corresponds to the upper sign choice. The quantities $f_1$ and  $f_2$ are
\begin{multline}
f_{1,2} \,=\,\frac{2}{\xi^2  \Delta_0} \big[ (\xi^2+1) \Delta_0 \mp 2 (c_n^2 + c_p^2)  (\xi^2+\xi) \sin^2(\theta) + 
\\
 \mp c_\varepsilon^2 (\xi^2+\xi+1) \cos(2 \theta) \pm c_\varepsilon^2 \xi \big]  
\, ,
\end{multline}
where $f_1$ corresponds to the upper sign choice and $\theta$ is the angle between $\boldsymbol{\Omega}$ and $\mathbf{k}$.
Finally, combining the above expressions, we find that the correction to the four branches of the dispersion relation for sound waves produced by the Coriolis force is
\beq
\omega(k)  
\, = \, 
\pm \left( c_{1,2} \, k \, + \, \dfrac{f_{1,2} \, \Omega^2}{2 \, c_{1,2} \, k} \right)  
\, .
\label{sound2}
\eeq
which, neglecting entrainment ($\varepsilon_\n=\varepsilon_\p=0$) reduces to \citep{waves}:
\beq
&&\omega \approx c_\n k\left[1+\frac{2 \Omega^2}{c_\n^2 k^2}(\sin\theta)^2\right] \\
&&\omega \approx c_\p k\left[1+\frac{2 \Omega^2}{c_\p^2 k^2}(\sin\theta)^2\right]\, .
\eeq

\subsection{Inertial waves}

We now continue to focus on the effect of rotation, but search for solutions that are inertial waves, for which the dispersion relation is not linear in $k$. If we set $c_\n=c_\p=C_\n=C_\p$ we find, as expected, families of inertial modes such that
\beq
&&\omega=\pm 2\Omega\cos\theta \\
&&\omega=\pm\frac{2\Omega \cos\theta}{(1-\varepsilon_{\n}-\varepsilon_{\p})}.
\eeq
In this case we still have two families of modes, but these do not correspond to oscillations in the single fluids as in the sound wave case, but rather represent an inertial mode where the two fluids flow together, which represents the standard inertial mode of the system, and a mode in which the two fluid oscillate relative to each other, the frequency of which is affected by entrainment \citep{rmode}.

\section{Including mutual friction}

So far it has been assumed that the two fluids can oscillate without any dissipation mechanism. The mutual friction will couple the two fluids and generally tend to damp any relative motion. To study this problem we consider a selection of analytically tractable cases, and defer the reader to section \ref{num_res} for the full analysis. We begin with a discussion of inertial waves, as it is known that inertial modes may be dynamically unstable in the presence of a background flow \citep{waves, Prix2004, Link2018}.

Let us start our analysis of inertial waves by considering the weak drag regime, such that $\mathcal{R}\ll 1$, which results in $\mathcal{B}^{'}\ll \mathcal{B}$. We thus take $\mathcal{B}^{'}=0$, and also set $c_\n=c_\p=C_\n=C_\p=0$ as we are interested in inertial waves. In the case in which there is no background flow, i.e. $w^i_{\n\p}=v^i_\n=0$, the dispersion relation follows from
\beq
&&\left[\left(4(\Omega\cos\theta)^{2}-\omega^2\right)\left(4(\Omega\cos\theta)^{2} \right.\right.\nonumber\\
&&\left.\left.-(2B {k}_i \Omega^i(1+\tilde\rho)-i(1-\varepsilon_\n-\varepsilon_\p)\omega\right)^2\right]=0 \, ,
\eeq
where $\tilde\rho=\rho_\n/\rho_\p$. The solutions are still an undamped inertial mode, which we identify with the co-moving mode, as in this case there is no relative motion and no mutual friction (note that in a full spherical analysis rotation couples the two motions at higher order, and results in mutual friction damping even for the standard co-moving mode, see \citealt{rmode}), and the damped counter-moving mode, modified by entrainment.
\be
\omega=\pm2\Omega\cos\theta
\ee
\beq
\omega =  \frac{2 \Omega\cos\theta}{1-\varepsilon_\n-\varepsilon_\p}(\pm 1- i \mathcal{B}(1+\tilde\rho))
\eeq
Note that the denominator is always positive, $1-\varepsilon_\n-\varepsilon_\p>0$, as required by stability on a microscopic scale \citep{CarterChamel2006}. The mode is therefore always stable, but we see here that our study of large scale hydrodynamical instabilities, such as the one that would arise if $1-\varepsilon_\n-\varepsilon_\p<0$, successfully capture a more general physical instability of the system.

\subsection{Background flow}

We now consider the physically realistic case in which the fluids are not co-rotating in the background, for example due to vortex pinning in the crust. The analysis is now more involved, so we will consider two cases separately: counterflow along the axis of the vortex array (the DG instability), and counterflow perpendicular to the axis. 

\subsubsection{The Donnelly-Glaberson instability}
\label{analyticDG}
To study the instabilities that arise due to counterflow along the vortex axis we specialize our setup, and consider only modes propagating along the vortex axis, which is taken to be alligned with the rotation axis along $z$. We thus have $k_x=k_y=0$. To begin our analysis we consider the simplified case with no entrainment, i.e. $\varepsilon_\x=\varepsilon_\y=0$ and also start from the assumption $\tilde{\rho} = \rho_n / \rho_p =0$, which corresponds to assuming that the mutual friction only acts on the neutrons. This not only simplifies the calculations, but it corresponds to assuming that the protons and electrons are 'clamped' to the normal fluid (e.g. their motion is entirely dictated by the magnetic field).  In this case the dispersion relation is:
\beq
&&\omega(k_z v_z^n+\omega)(4\Omega^2-\omega^2)((2\Omega-i\mathcal{B}k_z v_z^\n)^2\nonumber\\
&&-(2\mathcal{B}\Omega+i(k_z v_z^\n-\omega))^2)=0\, .
\eeq
Which, apart from the trivial solutions corresponding to the choice of coordinate system, has solutions:
\beq
&&\omega=\pm 2\Omega\\
&&\omega=\mp2\Omega-2 i \mathcal{B} \Omega \pm i B k_z v_z^n+k_z v_z^\n \, .
\eeq
The first solution corresponds to standard inertial waves (the factor $\cos\theta=1$ due to our choice of direction of $\mathbf{k}$), the second corresponds to the DG instability and is generally unstable as long as $k_z v^z>0$, i.e. $|v_\n|>2\Omega/|k_z|$. If we consider a standard pulsar such that $2\Omega\approx 100$ rad/sec and hydrodynamical scales such that $|k_z|<1$ cm$^{-1}$, then we have an instability if $|v_\n|>10^2$ cm/s, which corresponds to a lag of $\Delta\Omega\approx 10^{-4}$, in the outer layers of the star, which is easily sustainable by pinning forces in the crust \citep{Sevesopin}. Such an instability is thus always likely to be present, as vortex bending and large scale flows in the normal fluid, coupled to the superfluid, will always induce counterflow along the vortex array. The fact that the instability exists even if $\tilde{\rho}=0$ and mutual friction is not acting on the protons, suggests the instability will exist even in the presence of strong magnetic fields, as suggested by the analysis of \citet{Link2018}.

If we relax our approximations, and take $\tilde{\rho}\neq 0$ and include entrainment in our analysis the full solution is intractable. We thus consider the weak drag case, for which $\mathcal{R}\ll 1$, and we can take $\mathcal{B}^{'}\approx 0$. Furthermore we consider two limiting cases, the small entrainment limit, which is relevant for the core of the star, and the large entrainment limit, relevant for the crust. For small entrainment $\varepsilon_\n<\varepsilon_\p\ll 1$, we have
\beq
\left.
\begin{aligned}
\omega = &\pm 2 \Omega (1+\varepsilon_\p) +\\
&+\frac{4\mathcal{B}\varepsilon_\p \Omega^2(\mathcal{B}k_z v^z_\n \pm 2\mathcal{B}\Omega\mp i k_z v^z_\n)}{\mathcal{B}^2[(k_zv^z_\n)^2\pm4\Omega k_z v^z_\n+4\Omega^2]+ (k_zv^z_\n)^2}\, ,
\end{aligned}
\right.
\eeq
and
\beq
\left.
\begin{aligned}
\omega = &(1+\varepsilon_\n)(\pm2\Omega-2 i \mathcal{B} \Omega)\mp 2 i \mathcal{B} \Omega \varepsilon_\p - i \mathcal{B} k_z v_z^n +  \\
&+k_z v_z^\n-\frac{4\mathcal{B}\varepsilon_\p \Omega^2(\mathcal{B}k_z v^z_\n\pm 2\mathcal{B}\Omega \mp i k_z v^z_\n)}{\mathcal{B}^2[(k_zv^z_\n)^2\pm\Omega k_z v^z_\n+4\Omega^2]+ (k_zv^z_\n)^2}\, .
\end{aligned}
\right.
\eeq

We see that in the presence of a background flow entrainment modifies also the co-moving mode, which is now potentially unstable. For weak drag, if we expand to first order in $\mathcal{B}$, we see that the mode is always unstable for 
\be
\frac{1}{|k_z|}>\frac{|v_\n|}{4\mathcal{B}\varepsilon_\p\Omega^2}\, ,
\ee
which means that in the presence of a lag, due to core pinning (e.g. of neutron vortices to proton fluxtubes), the vortex array is generally unstable on all hydordynamical lengthscales for standard parameters ($\mathcal{B}\approx 10^{-4}$, $\Omega\approx 100$, $\varepsilon_\p=0.6$, $|v_\n|=10^4$ cm/s).

For the crust of the neutron star we can make the large entrainment approximation, we to simplify our calculation we take to be the limit $\varepsilon_\x\longrightarrow \infty$.
In this case the dispersion relation is:
\beq
\left.
\begin{aligned}
\omega=&-2\Omega(1+\tilde{\rho})+k_z v^z_\n-k_z v^z_\n(i \mathcal{B}-\tilde{\rho}(1+i \mathcal{B}))\\
& \pm i\sqrt{(1+\tilde{\rho})}S_1 \\
\omega=&+2\Omega(1+\tilde{\rho})+k_z v^z_\n+k_z v^z_\n(i \mathcal{B}-\tilde{\rho}(1+i \mathcal{B}))\\
& \pm i\sqrt{(1+\tilde{\rho})}S_2 \,
\end{aligned}
\right.
\eeq
with:
\beq
\left.
\begin{aligned}
S_1 = &\sqrt{8(1-i \B) k_z v^z_\n \Omega+(2i \Omega+(\mathcal{B}-i)k_z v^z_\n)^2} \\
S_2 = & \sqrt{-8(1-i B) k_z v^z_\n \Omega+(2i \Omega+(\mathcal{B}-i)k_z v^z_\n)^2} \, .
\end{aligned}
\right.
\eeq
from which we can see that one of the modes is always generically unstable in the limit $|k_z v^z_\n|/\Omega \ll 1$, which we expect to be satisfied in most cases in a NS interior. 

\subsubsection{Two stream instability} 
\label{AnalyticTS}
Let us now consider the case of a two stream instability, in which there is a counterflow perpendicular to the vortex axis, which we take to be aligned with the rotation axis such that $\hat{\kappa}^i=\hat{\Omega}^i=\hat{z}^i$. We thus have $v_{x}^\n \neq 0$, with $v_{y}^\n = 0, v_{z}^\n = 0$, and to keep the calculation tractable we consider the case in which $\varepsilon_\x=0$ and take the protons to be clamped to the normal fluid($\tilde {\rho}=0$). The dispersion relation is:
\beq
&&\omega(k_x v^x_\n-\B k_y v^x_\n-\omega)(4 \Omega^2 - \omega^2) \nonumber\\
&&(4(1+\B^2)\Omega^2-2i\B\Omega(-k_x v^x_\n+B\ k_y v^x_\n +2 \omega)- \nonumber \\
&&(k_x v^x_\n - \omega)(k_x v^x_\n - \B k_y v^x_\n - \omega))=0 \, .
\eeq
Apart from the trivial solution we now have solutions of the form
\beq
\omega &=& 2\Omega-2 i \B \Omega +k_x v^x_\n \pm \frac{1}{2} \B k_y v^x_\n \pm \nonumber \\
&&\frac{1}{2}\sqrt{1 -  i \frac{\B k_x v^x_\n}{2\Omega} + \frac{\B^2 k_y^2 (v_x^\n)^2}{16\Omega^2}}
\eeq
This mode can be unstable due to the induced flow in the $y$ direction caused by the background flow in the $x$ direction. An expansion for $\mathcal{B}\ll1$ reveals that the criterion for instability is:
\beq
-\text{Im}(\omega) = 2 \B \Omega + \frac{1}{2}\B k_x v^x_\n<0
\eeq
This gives the condition for instability to develop:
\beq
&&v_x^\n>\frac{4 \Omega}{|k_x|}, k_x>0 \\
&&v_x^\n<-\frac{4 \Omega}{|k_x|}, k_y<0
\eeq
Note, this condition doesn't contain dependence on $k_y$. To obtain this we need expand to the second order in small parameter $|v_\n|/\Omega$, which leads to:
\beq
-\text{Im}(\omega) = 2 \B \Omega + \frac{\B}{2}k_x v^x_\n - \frac{\B^3 k_x^3 (v^x_\n)^3}{64 \Omega^2} - \frac{\B^3 k_x k_y^2 (v^x_\n)^3}{64\Omega^2}  \, .
\eeq
As we can see mutual friction doesn't affect the critical velocity for the instability to set in,  but affects its growth rate, with higher mutual friction causing a faster rise of the instability with increasing velocity.

\subsection{The strong drag regime}

We now move on to consider the case of a strong drag, that corresponds to $\R\gg 1$, in which case the parameter $\B^{'}$ can no longer be neglected. This case may be applicable in regions with strong pinning \citep{s3,s1,s2}, and may have important consequences for the development of the r-mode instability \citep{rmode, Trigger09}.
Let us begin our analysis by the equivalent of the DG instability, in which the relative flow is along the vortex axis, taken to coincide with rotational axis, or $v_{x}^{\n}=v_{y}^{\n}=0$ and $v_{z}^{\n} \neq 0$, and we also take $\varepsilon_\x=0$. We consider the limit $\mathcal{R}\ll 1$ for which $\B\approx 0$ and $\B^{'}\approx 1$. In this case the spectrum is given by:

\beq
\left.
\begin{aligned}
&\omega = 2\Omega-\B^{'}\Omega(1+\tilde{\rho})-\frac{1}{2} k_z v^z_\n +\frac{1}{2} B' k_z v_z^n \pm \frac{1}{2}\sqrt{T_1}\\
&\omega=-2\Omega+\B^{'}\Omega(1+\tilde{\rho})+\frac{1}{2}k_z v^z_\n +\frac{1}{2} B' k_z v_z^n \pm \frac{1}{2}\sqrt{T_2}\, ,
\end{aligned}
\right.
 \label {BTF}
\eeq
with
\beq
\left.
\begin{aligned}
T_1=&\B^{'2}(-2\Omega(1+\tilde{\rho})+k_z v^z_\n)^2\\
&+2 \B^{'} k_z v^z_\n [2\Omega(-1+\tilde{\rho})+k_z v^z_\n)]+(k_z v^z_\n)^2\\
T_2=&\B^{'2}(2\Omega(1+\tilde{\rho})+k_z v^z_\n)^2\\
&+2 \B^{'} k_z v^z_\n [2\Omega(1-\tilde{\rho})+k_z v^z_\n] +(k_z v^z_\n)^2\, .
\end{aligned}
\right.
\eeq
For standard neutron star parameters these modes are not unstable, and in fact if we expanding far $B^{'}\ll1$ we obtain:
\beq
&&\omega = 2 \Omega +2 \B^{'} \Omega +k_z v^z_\n \pm \B^{'} k_z v^z_\n \nonumber \\
&&\omega = - 2 \Omega + 2 \B^{'} \Omega + k_z v^z_\n \pm \B^{'} k_z v^z_\n
\eeq
Which are stable, modified inertial modes. 

\subsubsection{Sound waves and mutual friction}

To study the effect of mutual friction on sound waves we can consider an expansion in $\Omega$ around the $\Omega=0$ solution  $\omega= \pm c_\n k$. Let us consider the modes of the superfluid, in the clamped proton approximation. Defining $\theta$ to be the angle between the wave vector and the rotation axis, we find, for $\varepsilon_\x=C_\x=w_{\x\y}=v_\n=0$, we have:
\begin{multline}
\omega(k) 
\, = \, 
\pm c_\n k - i \B \Omega^2  \sin^2  \theta +
\\
\pm \dfrac{\Omega^2 \, \sin^2 \theta }{2 c_\n k}(4-8\B^{'}+4\B^{' 2}-5\B+\B^2 \cos2\theta )
\, .
\end{multline}
The effect of the  mutual friction disappears when the wave vector is aligned with the vorticity, i.e. $\theta=0$: in this case the velocity perturbation is parallel to the vortex array and there cannot be any mutual friction effect (all the cross products in \ref{MFD} are zero). Things are very different if a background velocity lag is present (even if this lag is parallel to the vortex array) because vorticity perturbations enter the game: this will be investigated numerically in the next sections.

In the weak drag limit, we may set $B'\approx \mathcal{R}^2$ and $\B\approx \mathcal{R}$ and the relevant terms become
\begin{multline}
\omega 
\, = \, 
\pm c_\n k - i \mathcal{R} \Omega^2 \sin^2  \theta 
\pm \dfrac{ \Omega^2 \sin^2 \theta}{2 c_\n k}(4-5\mathcal{R} )
\, .
\end{multline}

\subsection{Isotropic (Gorter-Mellink) mutual friction}
\label{GMfriction}
We now consider the Gorter-Mellink form for the mutual friction in (\ref{GM}) to investigate if and how the presence of an isotropic vortex tangle modifies the previous results. 
In the limit $k c_\n \gg A_{GM} |\mathbf{v}_n|^2$ it is possible to write the exact implicit form of the dispersion relation, which is: 
\be
\left.
\begin{aligned}
\omega(k)\, & = \, \pm 2 \Omega |\cos \theta| -i \alpha_I^{\pm}  A_{GM} |v_\n|^2 \left( 1- \dfrac{f_I^{\pm} \Omega^2}{c_\n k^2} \right)
\\
\omega(k)\, & = \, \pm \left( c_\n k+\dfrac{2 \Omega^2 }{ c_\n k^2} \right) - i \alpha_S^{\pm}  A_{GM} |v_\n|^2 \left(1- \dfrac{ f_S^{\pm} \Omega^2}{c_\n k^2} \right)
\end{aligned}
\right.
\label{GMmodes}
\ee
for inertial and sound waves respectively. Here $f^{\pm}_I\sim 1$ and $f^{\pm}_S\sim 1$ are four involved functions of the angles between the vectors $\mathbf{v}_n$, $\mathbf{k}$ and $\boldsymbol{\Omega}$, as well as the two positive functions $\alpha^\pm_I> 0$ and $\alpha^\pm_S>0 $. Irrespectively of the mutual orientation and magnitude of the three vectors (provided that $k c_n$ is always much bigger than $\Omega$ and $A_{GM} |\mathbf{v}_n|^2$), we are always in a damped regime in which the damping timescale is of the order of $A_{GM}^{-1} |\mathbf{v}_n|^{-2}$ for both inertial and sound waves.

It would thus appear that a rectilinear vortex array is rapidly destabilized as unstable inertial modes develop due to counterflow. If an (approximately) isotropic vortex tangle develops this remains stable, at least until the turbulent tangle decays. Once a rectilinear array is restored the system is again unstable, leading to a recurrent mechanisms that may be linked to the trigger of pulsar glitches. A more detailed analysis in the case of a polarized turbulent tangle should be the focus of future work.

Note that as the modes in (\ref{GMmodes}) are stable, we will not consider isotropic mutual friction in the numerical analysis in the following section.

\section{Numerical results}
\label{num_res}

\begin{figure*}

\begin{minipage}[b]{0.49\linewidth}
\centering
\includegraphics[width=.9\linewidth]{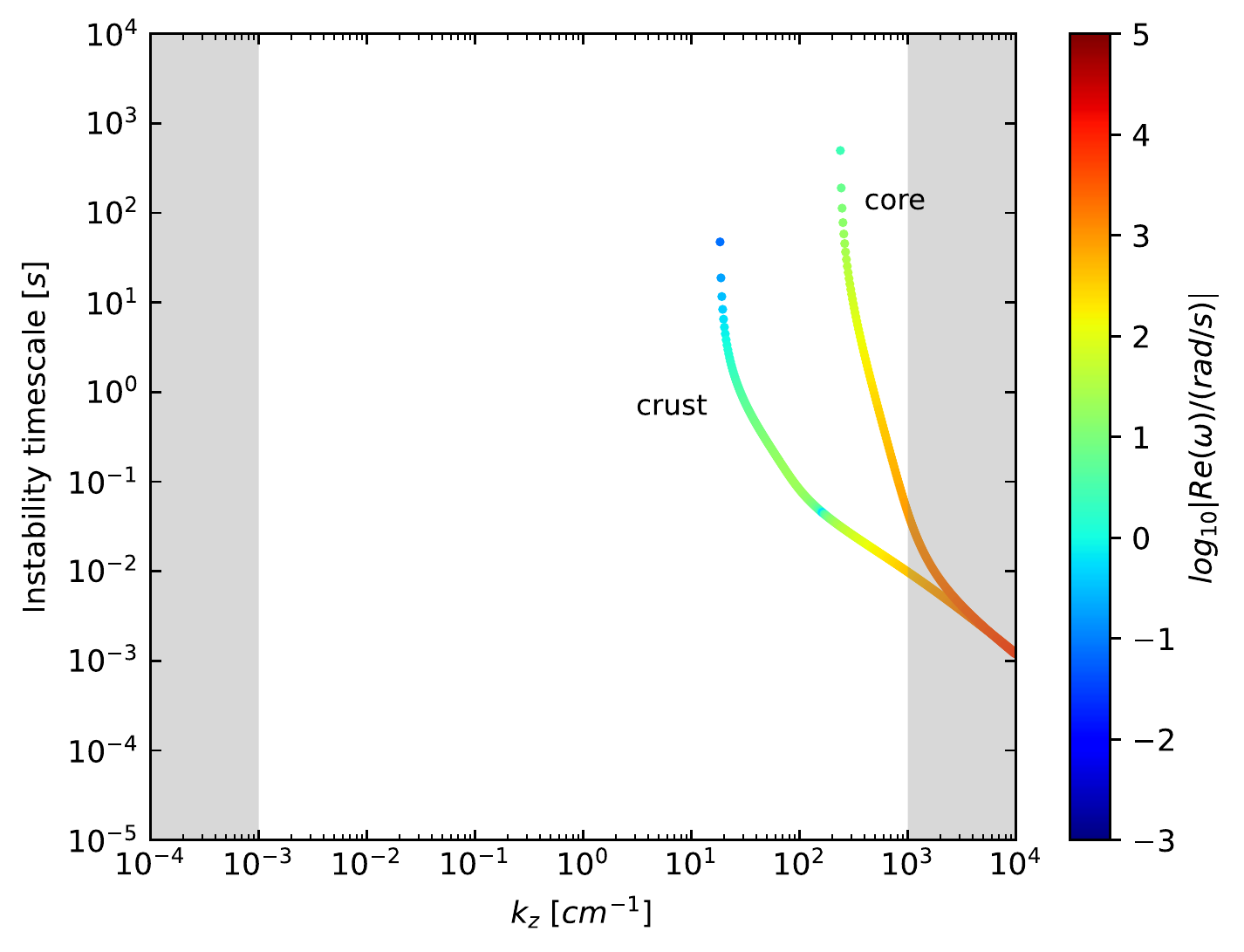} 
%\label{DG3}
%\label{fig:1vZkZ_1e0}
\vspace{4ex}
\end{minipage}
\begin{minipage}[b]{0.49\linewidth}
\centering
\includegraphics[width=.9\linewidth]{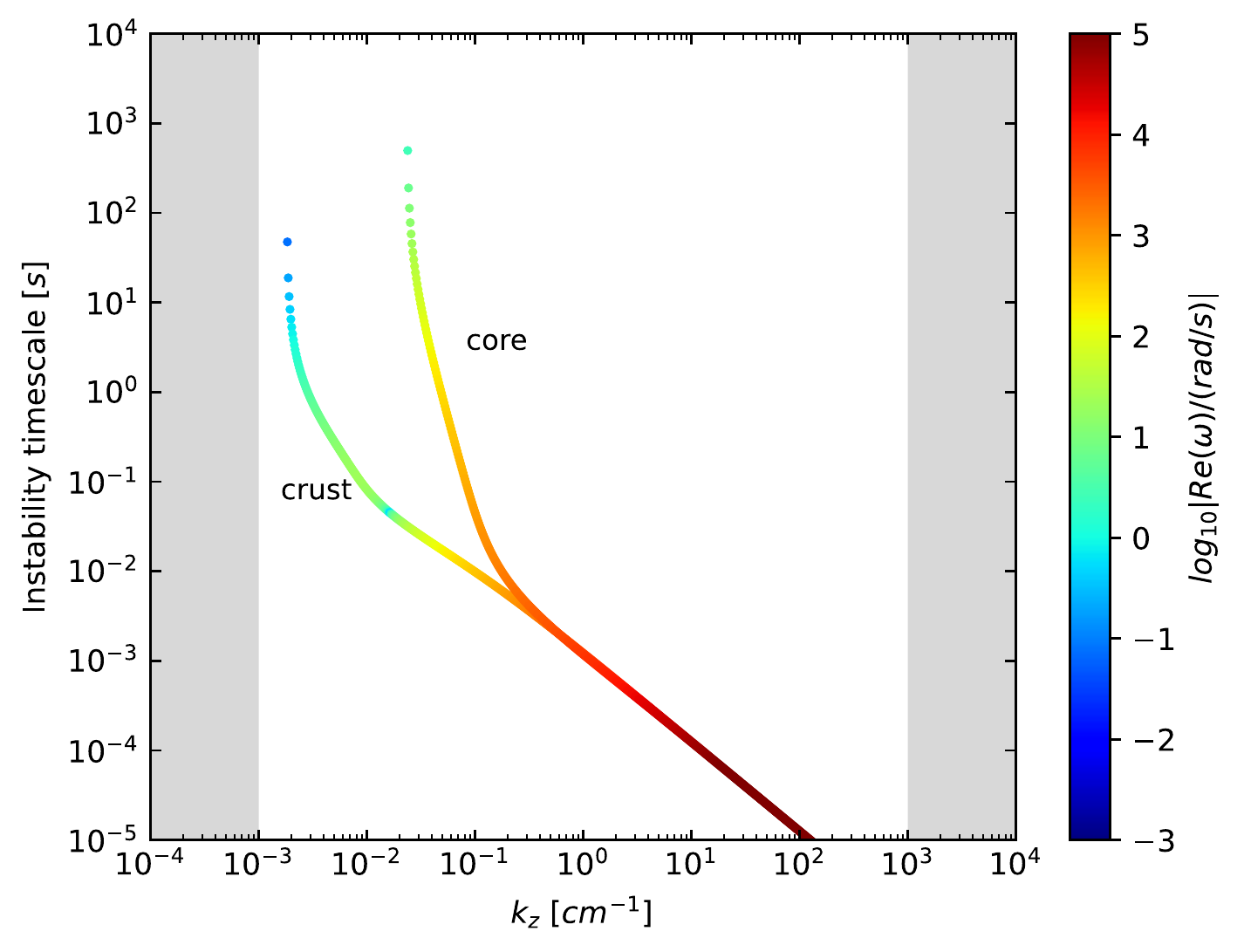}

%\label{fig:3vYkY_1e0}
\vspace{4ex}
\end{minipage} 

\caption{Instability timescale versus wave-vector $\mathbf{k}$ for the DG instability, in which the counterflow and $k$ are along the vortex axis in the $z$ direction. The colour coding expresses the real part of the oscillation frequency (note that the rotation rate of the star, i.e. of the `normal' component, is taken to be $\Omega=100$ rad/s). We consider the strong drag case with $\B=\B^{'}=0.5$ and two values for the background lag, a low value of $v_\n^z=1$ cm/s (left), and a high value (corresponding approximately to the maximum that pinning forces can sustain), of $v_\n^z=10^4$ cm/s (right). In general we see that there are always mixed inertial-sound waves, that are unstable on dynamical timescales, and are increasingly unstable on small lengthscales, and that for large enough background velocity lags, the whole dynamical range is unstable.}
\label{Fig:DG3}
\end{figure*}
\begin{figure*}

\begin{minipage}[b]{0.49\linewidth}
\centering
\includegraphics[width=.9\linewidth]{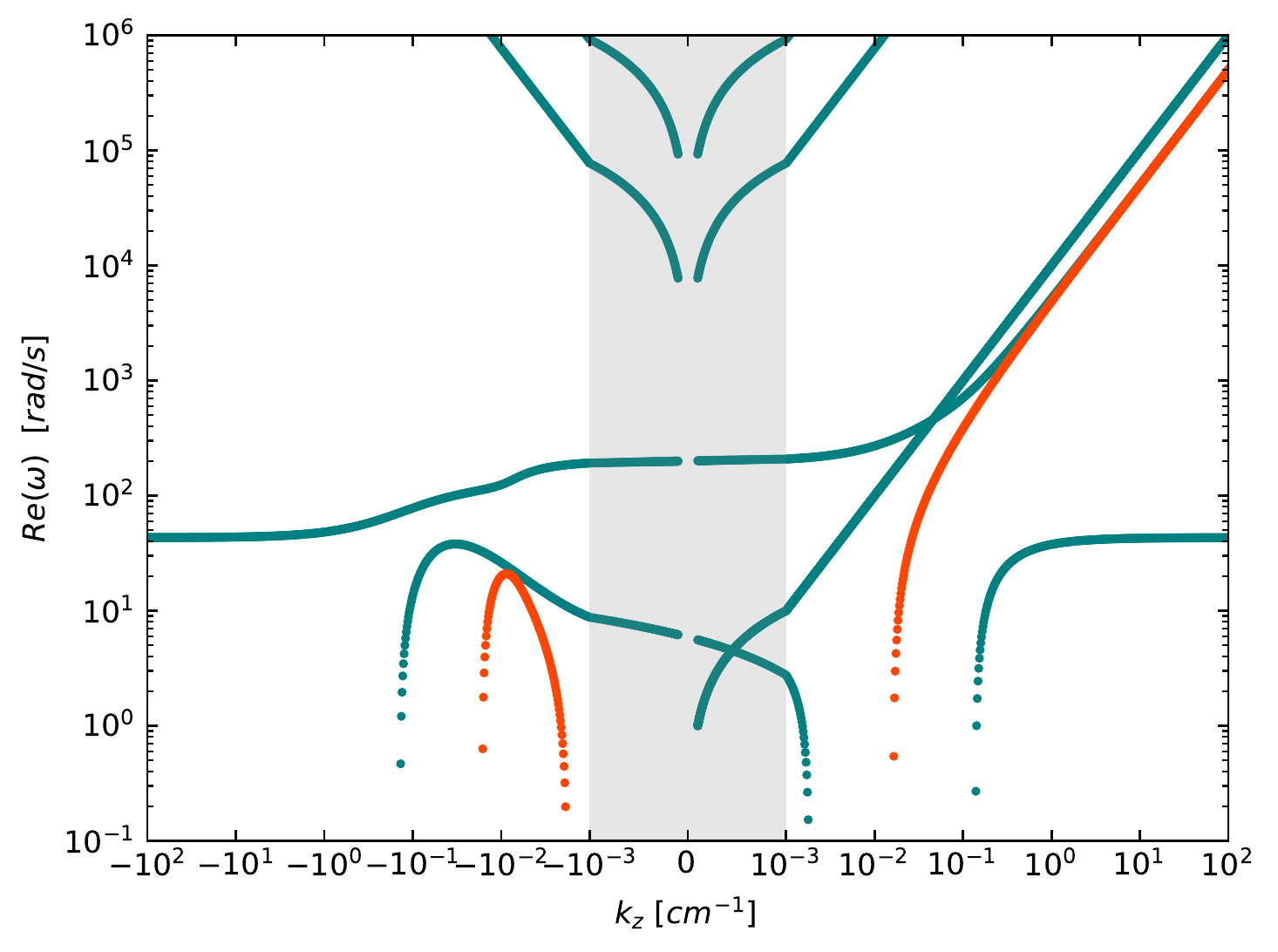} 
%\label{DG3}
%\label{fig:1vZkZ_1e0}
\vspace{4ex}
\end{minipage}
\begin{minipage}[b]{0.49\linewidth}
\centering
\includegraphics[width=.9\linewidth]{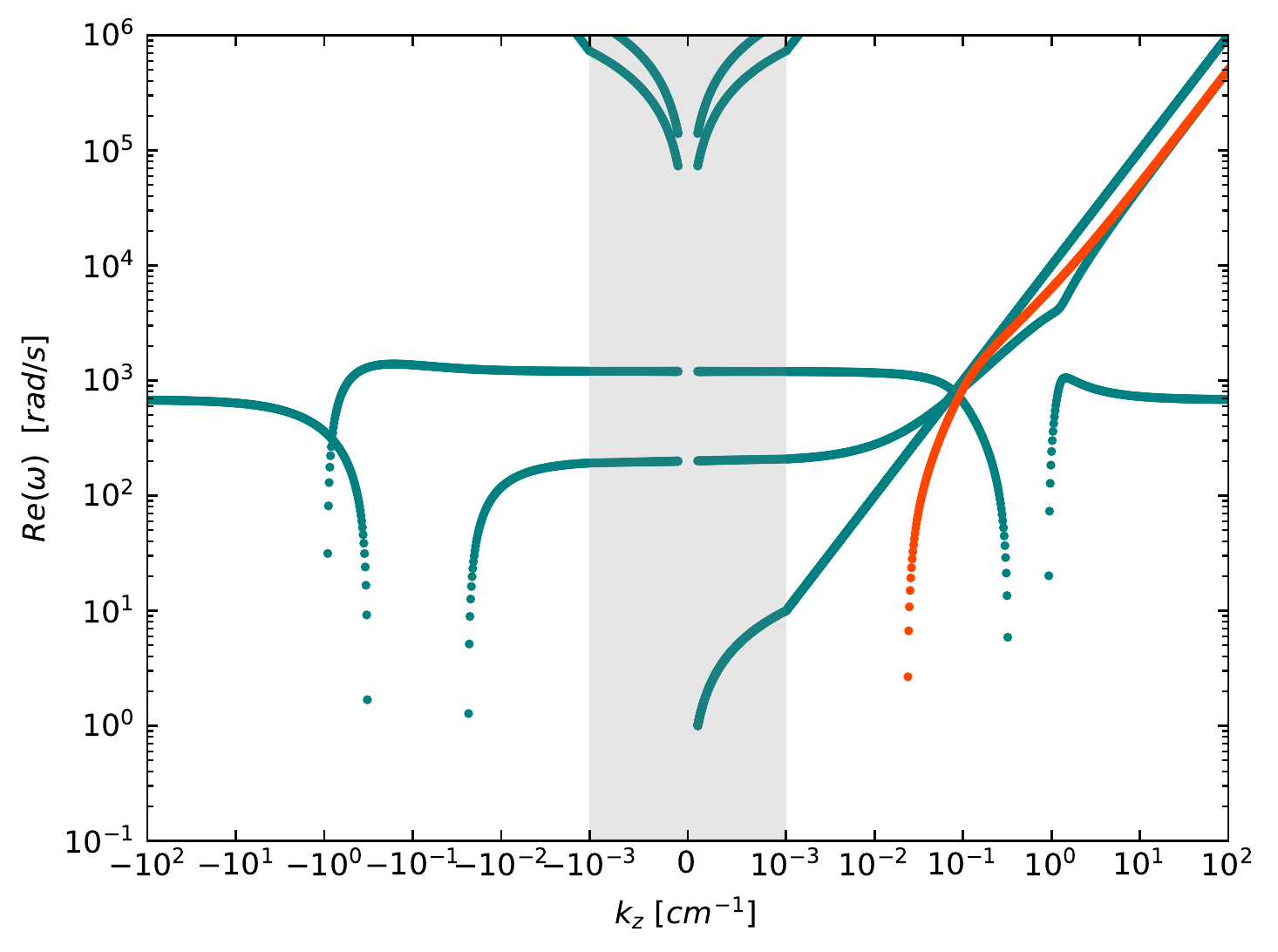}

%\label{fig:3vYkY_1e0}
\vspace{4ex}
\end{minipage} 

\caption{Dispersion relations (Real part of $\omega$ vs $k_z$) for the crust (left) and core (right) of the neutron star, for a lag of $v_\n^z=10^4$ cm/s in the case of strong drag ($\mathcal{B}=\mathcal{B}^{'}=0.5$), for the DG instability in which we consider the background lag along the vortex axis, i.e. the $z$ axis. We can see that there are mode crossings between families of inertial and sound waves, and in red we have the unstable modes, which are modified sound-inertial waves. In the crust, in the presence of large entrainment, an additional unstable inertial mode is present for negative values of $k_z$. The shading identifies the region in which the scale of the horizontal axis is linear.}
\label{Fig:Disperdo1}
\end{figure*}   
   
All the machinery needed to derive the general dispersion relations for the various modes of oscillation of the two-fluid system is presented in  Appendix \ref{appendixequations}. 
The final result is the matrix $M$ in \eqref{Mfinal}, which depends on some basic local quantities that define how the system responds to a perturbation in the velocity fields: the full matrix (or its determinant) can be seen as a function of $\omega$ and $\mathbf{k}$ and depends on several parameters, namely 
\beq
M  =  M( \omega, \mathbf{k} \, ; \, \mathbf{v}_{n}, \boldsymbol{\omega}, \boldsymbol{\Omega} \, ; \, \varepsilon_\x, \rho_\x, \alpha_\x, A_\x, C_\x, c_\x \, ; \, \mathcal{B} , \mathcal{B}' \, )  .
\label{M66}
\eeq
Leaving aside the obvious dependence on the pulsation $\omega$ and on $\mathbf{k}$, the other parameters (that describe completely the hydrodynamic state of the background configuration when there is no turbulence and magnetic field) have been formally divided into three sets, which helps us to discuss their role. 
\\
\\
The first set comprises the variables $ \mathbf{v}_{n}$, $\boldsymbol{\omega}$ and $\boldsymbol{\Omega}$: they define the state of motion of the background configuration in which the normal component rotates rigidly. We allow for a stationary non-zero local velocity lag $\mathbf{v}_{n}=\mathbf{w}_{\n\p}$, which is expected to be non constant on the stellar radius length scale. 
Therefore, in a purely local analysis, the background vorticity field $\boldsymbol{\omega}$ is not expressible in terms of the local (constant) value of  $\mathbf{w}_{np}$: only a true solution of the unperturbed equations of motion would lead to locally consistent values for $ \mathbf{w}_{\n\p}$, $\boldsymbol{\omega}$ and $\boldsymbol{\Omega}$, but would prevent us to test the more general case in which the local direction and magnitude of these vectors are chosen at will. 
\\
\\
The second set of parameters describes the local state of matter in the stationary configuration. It can be further divided into two subsets. The four parameters $\epsx$ and $\rho_\x$, are defined by considering the first law of thermodynamics \eqref{first_thermo} and their unperturbed value can oscillate: indeed, we introduced the linear combinations \eqref{rho_amplitude} and \eqref{amp_entr} to express their amplitudes in terms of the perturbed velocities (the same is valid for the chemical potentials $\tilde{\mu}_\x$, with the only caveat that the  unperturbed value of the chemical potentials do not enter into the explicit expression of $M$). 
The second subset comprises $\alpha_\x$, $A_\x$, $C_\x$ and $c_\x$, that are related to second order derivatives of the internal energy $E$: in a first-order analysis these quantities are fixed to their unperturbed value and allow, together with the unperturbed values $\rho_\x$ and $\epsx$, to express the fundamental thermodynamic perturbations $\delta\rho_\x$,  $\delta\epsx$ and $\delta\tilde{\mu}_\x$  as  described in section \ref{termo}.
\\
\\
Finally, the two parameters $\mathcal{B}$ and $\mathcal{B}'$ define the ``state'' of the vortex array (note again, that we only consider the standard, anisotropic form of the mutual friction, and not the Gorter Mellink form, as we have found modes to be stable for this form of the mutual friction). We do not consider variations of these parameters and their value has to be fixed according to the mutual friction scenario that we are interested to test. In the subsequent numerical analysis  we consider different possibilities, listed in table \ref{tab:tab2}.
\\
\\

\begin{figure*}
\begin{minipage}[b]{0.49\linewidth}
\centering
\includegraphics[width=.9\linewidth]{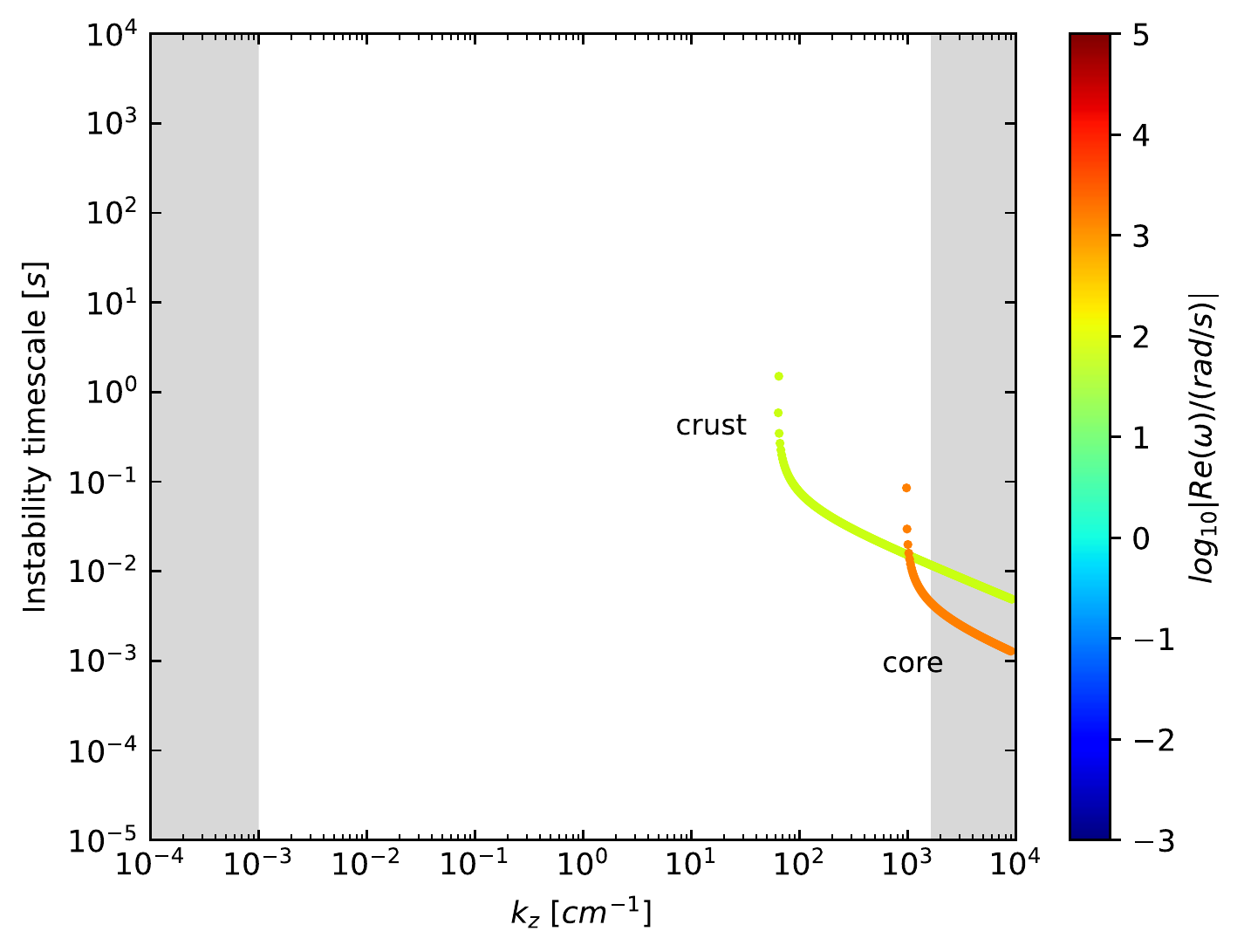}

%\label{fig:3vZkZ_1e0}
\vspace{4ex}
\end{minipage}
\begin{minipage}[b]{0.49\linewidth}
\centering
\includegraphics[width=.9\linewidth]{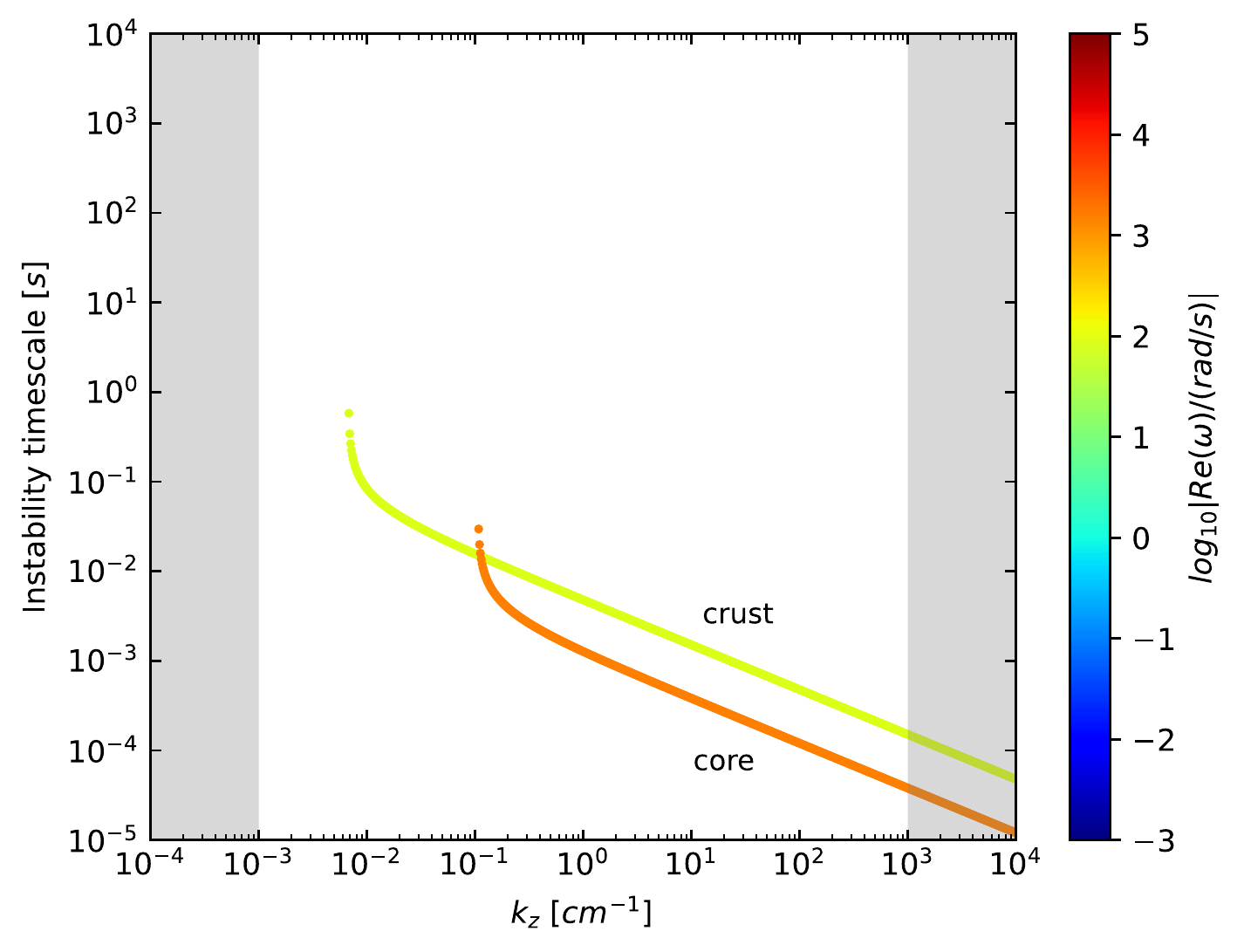}
\vspace{4ex}
\end{minipage} 

\caption{Instability timescale versus wave-vector $\mathbf{k}$ for the DG instability, in which the counterflow and $k$ are along the vortex axis in the $z$ direction. The colour coding expresses the real part of the oscillation frequency (note that the rotation rate of the star, i.e. of the `normal' component, is taken to be $\Omega=100$ rad/s). The setup us the same as in Figure \ref{Fig:DG3}, but here we consider the `pinned' case with $\B=0$ and $\B^{'}=1$, again two values for the background lag, a low value of $v_\n^z=1$ cm/s (left), and a high value of $v_\n^z=10^4$ cm/s (right). We see that we still have unstable modes, although if one observes the dispersion relation in figure \ref{Fig:Disperdo2}, it is clear that these are now inertial waves and not mixed sound-inertial waves as in the strong drag case (the weak drag case is qualitatively similar to the strong drag one).}
\label{Fig:DG4}
\end{figure*}

\begin{figure*}

\begin{minipage}[b]{0.49\linewidth}
\centering
\includegraphics[width=.9\linewidth]{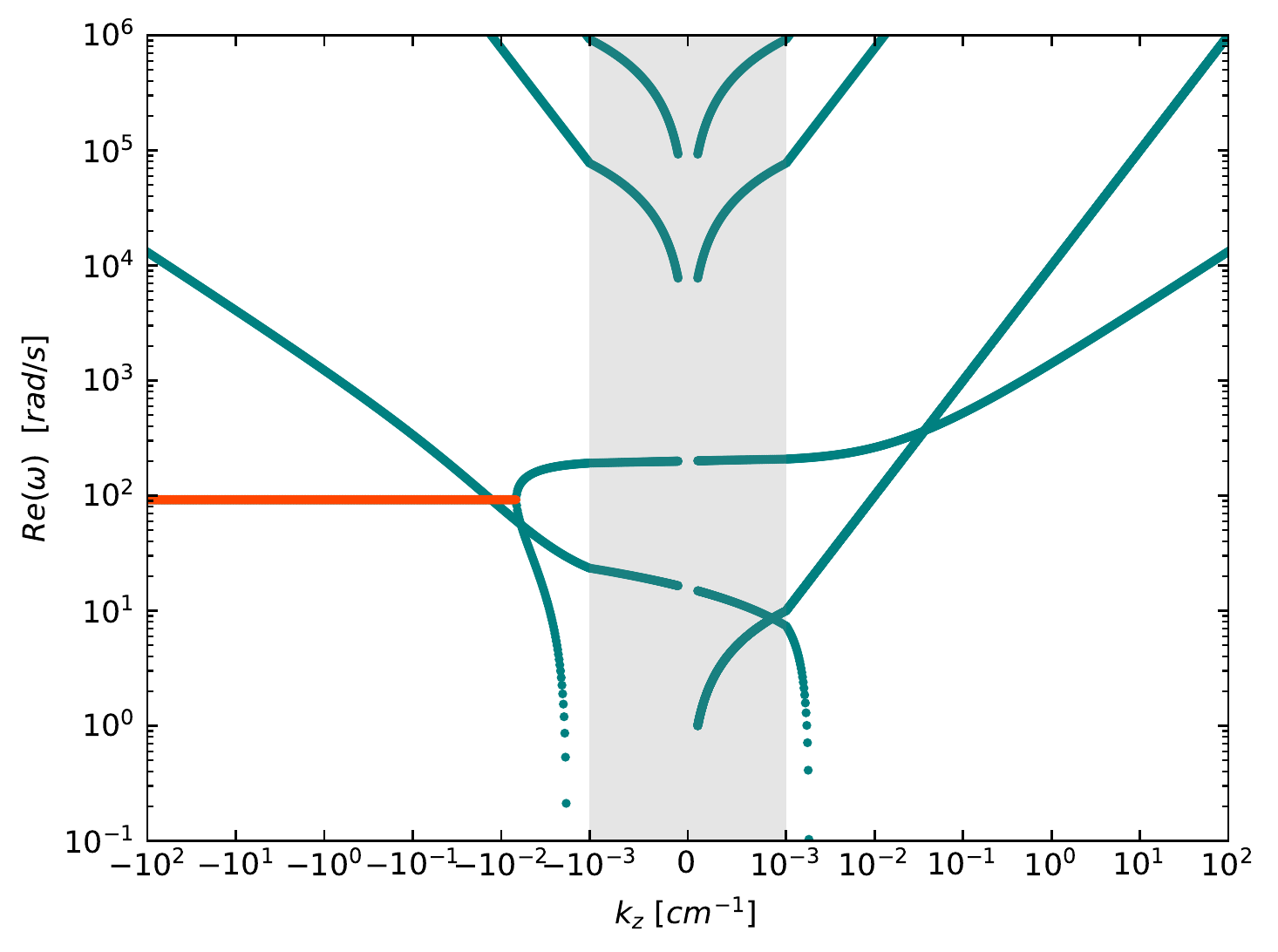} 
%\label{DG3}
%\label{fig:1vZkZ_1e0}
\vspace{4ex}
\end{minipage}
\begin{minipage}[b]{0.49\linewidth}
\centering
\includegraphics[width=.9\linewidth]{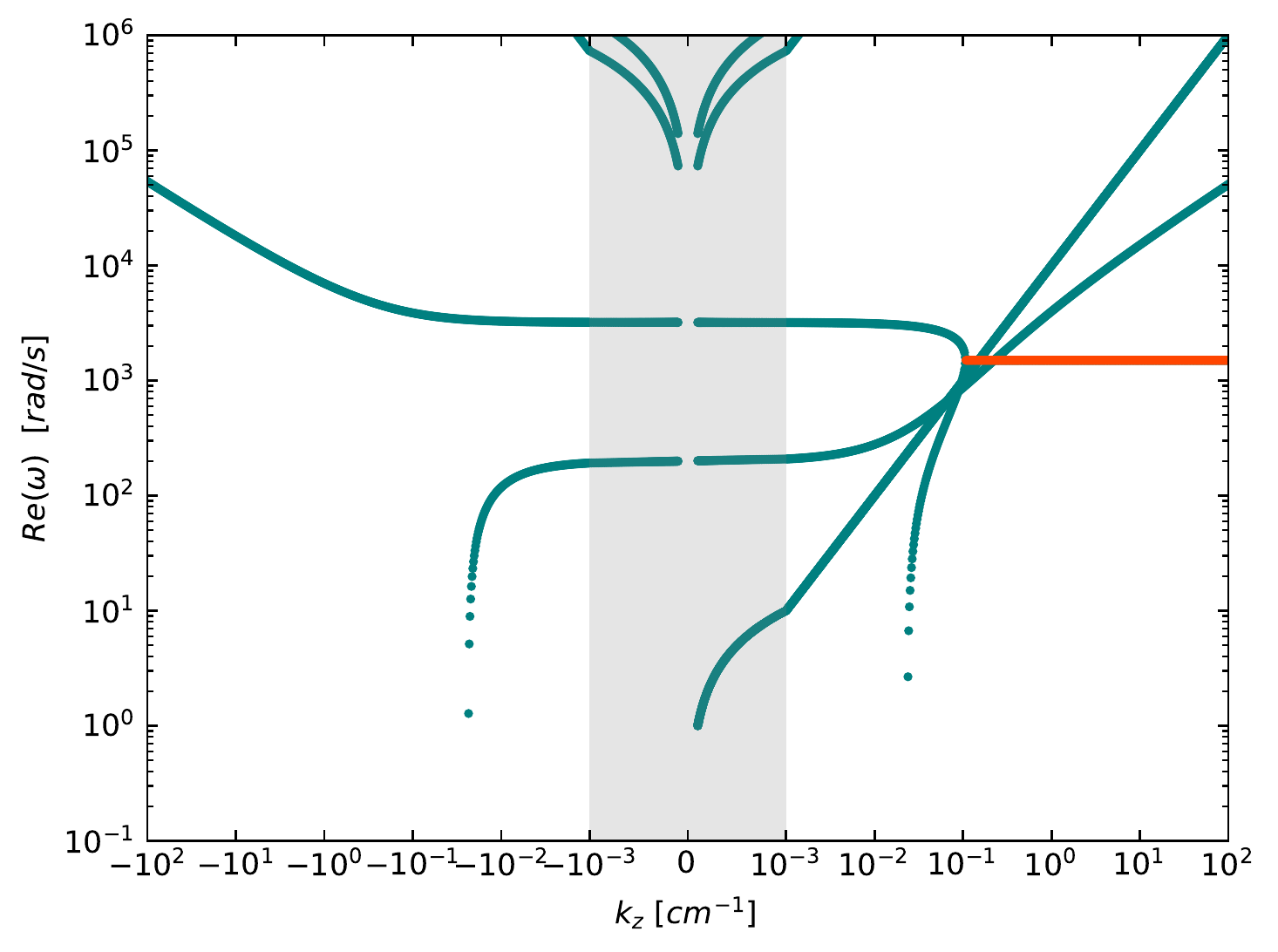}

%\label{fig:3vYkY_1e0}
\vspace{4ex}
\end{minipage} 

\caption{Dispersion relations (Real part of $\omega$ vs $k_z$) for the crust (left) and core (right) of the neutron star, for a lag of $v_\n^z=10^4$ cm/s for the pinned case ($\mathcal{B}=0, \mathcal{B}^{'}=1$), in the case of the DG instability in which we consider the background lag along the vortex axis, i.e. the $z$ axis. Unlike in the strong drag case in Figure \ref{Fig:Disperdo1},  the unstable modes (in red) are now clearly inertial waves. The shading identifies the region in which the scale of the horizontal axis is linear.}
\label{Fig:Disperdo2}
\end{figure*}   

From the analytic point of view, writing down the full matrix $M$ is of little interest but its complete and explicit form in components can be easily obtained with the aid of any software for symbolic computation, as outlined in appendix \ref{appendixequations}. Although extremely complex, also the full and exact determinant of $M$ can be obtained as well. At this point we substitute a particular choice for the background parameters into the complete expression of the determinant; the roots of $\det M(\omega, \mathbf{k}) =0$, which is a high degree polynomial equation in the pulsation $\omega$, can be computed numerically for different values of $\mathbf{k}$. By varying $\mathbf{k}$ and the parameters, some branches of the dispersion relation $\omega(\mathbf{k})$ may result in a positive imaginary part. The associated instability timescale is then defined as 
\beq
T(\mathbf{k}) \, = \, \dfrac{2 \, \pi}{\text{Im}(\omega(\mathbf{k})) } \, .
\label{timescale}
\eeq
Since we worked within a purely hydrodynamic framework, it is important to stress that not all the values of $|\mathbf{k}|$ are physically meaningful. As also discussed in \cite{Link2018}, the wave vector should be much smaller than the inverse of the stellar radius, otherwise the effect of stratification is expected to modify the dispersion relation (namely, the background quantities have spatial dependence on such length scales and cannot be considered uniform, as in the present local analysis). Since the stellar radius is about $R \sim 10\,$km, we should consider   $|\mathbf{k}| \gg 10^{-6}\,$cm$^{-1}$: therefore, we expect the present analysis to be valid for  $|\mathbf{k}| \sim 10^{-3}\,$cm$^{-1}$ since significant density changes are expected to occur on the length scale of about ten meters, especially at the core-crust interface. 

On the other hand, the definition of the superfluid momentum and vorticity needs a suitable average on a macroscopic sample of matter containing many vortex lines, which average separation is expected to be of the order of $\sim 10^{-3}\,$cm. Therefore, we expect the present analysis to break down when the wave vector reaches the critical value $|\mathbf{k}| \sim 10^{3}\,$cm$^{-1}$.
For this reason,    the region outside the physically interesting range $10^{-3} < |\mathbf{k}|\,$cm$\,< 10^{3}$ is shaded in   the plots of the instability timescales.

\begin{table*}
\centering
\begin{tabular}{ccccccccccc}
  \hline 
  Case & $\rho_n$ & $c_n$ & $c_p$ & $\varepsilon_n$ & $\varepsilon_p$ & $\Omega$  & $|\boldsymbol{\omega}|$  & $ \hat{\boldsymbol{\omega}} $ & $A_{n,p}$ & $C_{n,p}$ \\ 
  \hline 
  \hline 
  Crust & $4\rho_p$ & $10^9\,$cm/s & $0.51c_n$ & -10 & -40 & $100\,$rad/s & $200\,$rad/s & $\hat{\mathbf{z}}$ & 0 & 0 \\ 
  %\hline 
  %\hline 
  Core & $4\rho_p$ & $10^9\,$cm/s & $0.51c_n$ & 0.15 & 0.6 & $100\,$rad/s & $200\,$rad/s & 
  $\hat{\mathbf{z}}$ & 0 & 0 \\ 
  %\hline 
\end{tabular}   
  \caption{The two prototype cases, for the crust and core of the star, that have been tested in the numerical analysis of the determinant. For each of these five cases we consider nine different relative orientations of the local lag $\mathbf{v}_n$ and $\mathbf{k}$ and the three mutual friction scenarios listed in table \ref{tab:tab2}. Each of these  cases has been investigated by considering two relative velocity speeds between neutrons and the normal component: $|\mathbf{v}_n|=1 \,$cm/s and $|\mathbf{v}_n|=10^4 \,$cm/s }
  \label{tab:tab1}
\end{table*}

\begin{table}
\centering
\begin{tabular}{cccc}
  \hline 
  Description &$B'$ & $B$ & $\R$\\ 
  \hline 
  \hline 
Pinned vortices & 1 & 0  & $\infty$ \\ 
  %\hline 
Free vortices &  0 & 0 & 0\\ 
  %\hline 
Strong drag & 0.5 & 0.5 & 1 \\ 
  %\hline 
Weak drag & $10^{-4}$ & $10^{-2} $ & $\approx 10^{-2}$\\ 
  %\hline 
\end{tabular}   
  \caption{The mutual friction scenarios considered in the numerical analysis: for each case listed in table \ref{tab:tab1} we consider these four mutual friction regimes. }
  \label{tab:tab2}
\end{table}
  
Considering the relative orientation of four different vectors ($\mathbf{v}_n$, $\boldsymbol{\Omega}$, $\boldsymbol{\omega}$ and $\mathbf{k}$) leads to a huge parameter space; therefore, in the numerical analysis we stick to the case in which $\boldsymbol{\Omega}=\Omega \hat{\mathbf{z}}$ and $\boldsymbol{\omega}=2 \boldsymbol{\Omega}$, while $\mathbf{v}_n$ and $\mathbf{k}$ are chosen to be aligned with one of the three directions $\hat{\mathbf{x}}$, $\hat{\mathbf{y}}$ or $\hat{\mathbf{z}}$, for a total of nine combinations\footnote{
The presence of a non zero background lag may locally modify the vorticity direction. However, in our local analysis the background lag is treated as locally uniform, while the local direction of the vorticity depends on the large scale, global arrangement of the background lag in the stationary configuration: we are therefore forced, due to the fact that the analysis is local, to assume three  independent directions for the wave vector, the velocity lag and the vorticity. However, in the following analysis, we keep the vorticity fixed along the z-axis (i.e. aligned with $\mathbf{\Omega}$). This assumption, as well as the fact that we will use $|\mathbf{\omega}|=2 \Omega$ in the background configuration, is not severe when the lag in angular velocity is small compared to the rotation rate of the star. Note that only the background vorticity is fixed, we account for variations in  $|\boldsymbol{\omega}|$ in the mutual friction.
}. 
However, for simplicity, only the interesting cases that allow for unstable modes are shown in figures and discussed.
 
%Note that, despite $\boldsymbol{\omega}=2\boldsymbol{\Omega}$ is the usual result that one would use in the case of a rigidly corotating background configuration, there is no inconsistency in doing this: since our analysis is purely local, the background vorticity should be settled at will, or calculated consistently on the real background configuration. 
%We do not expect, however, large departures from the corotating case, at least as long as turbulence dis not present in the system. Departures of $|\boldsymbol{\omega}|$ from the corotation value due to the presence of a lag in the background configuration are expected to be small and have the only effect to change slightly the assumed values of $B'$ and $B$.  
% 
% 
%% LIMITATIONS: the star is unmagnetized and the analysis is not valid near the rotation axis (because we considered a local Cartesian set of coordinates aligned with the global cylindrical coordinates, therefore $\nabla_x$ is not simply $\partial_x$. However, the choice of Cartesian coordinates is coherent with the plane wave choice. 

\begin{figure*}
\begin{minipage}[b]{0.49\linewidth}
\centering
\includegraphics[width=.95\linewidth]{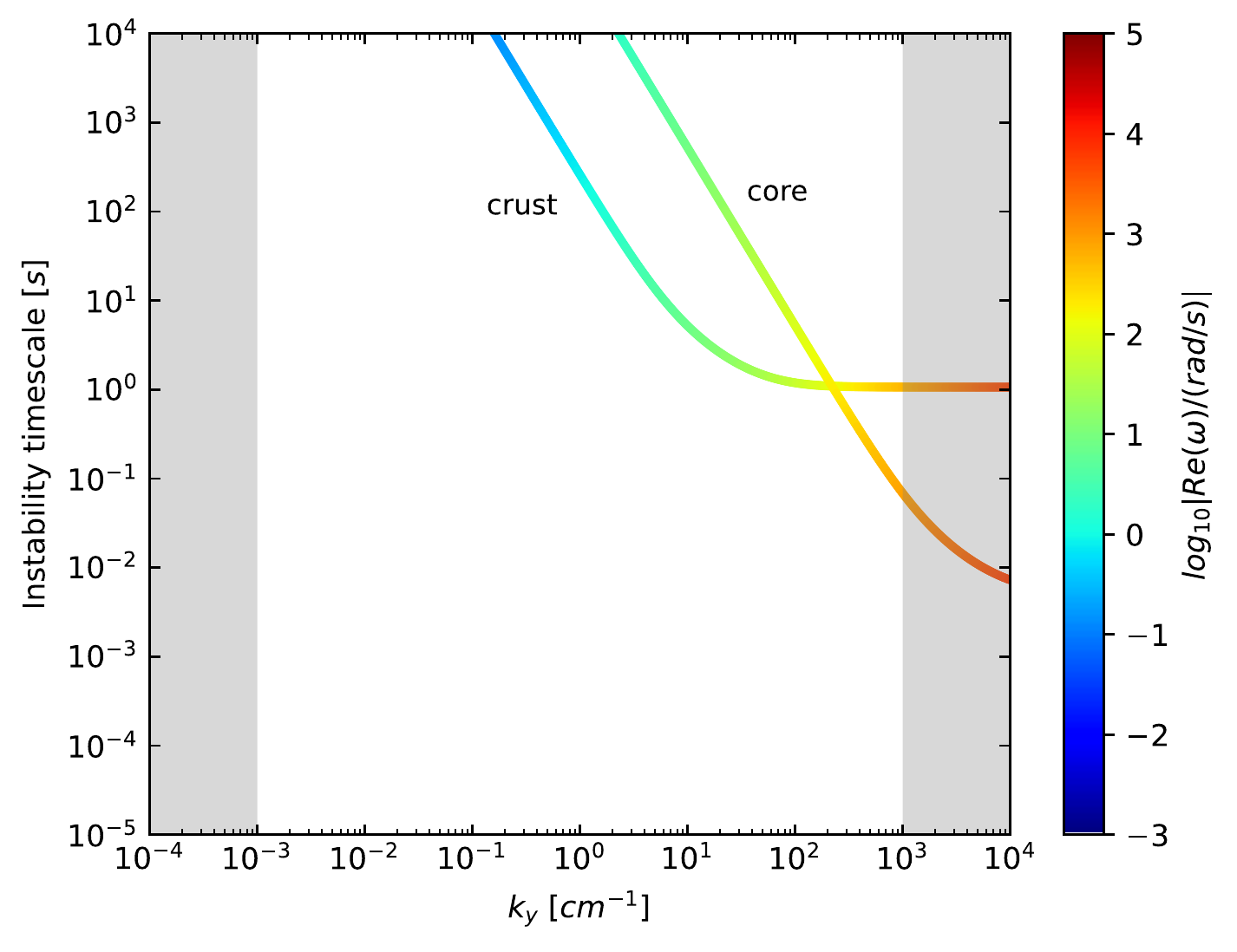}

%\label{fig:3vZkZ_1e0}
\vspace{4ex}
\end{minipage}
\begin{minipage}[b]{0.49\linewidth}
\centering
\includegraphics[width=.95\linewidth]{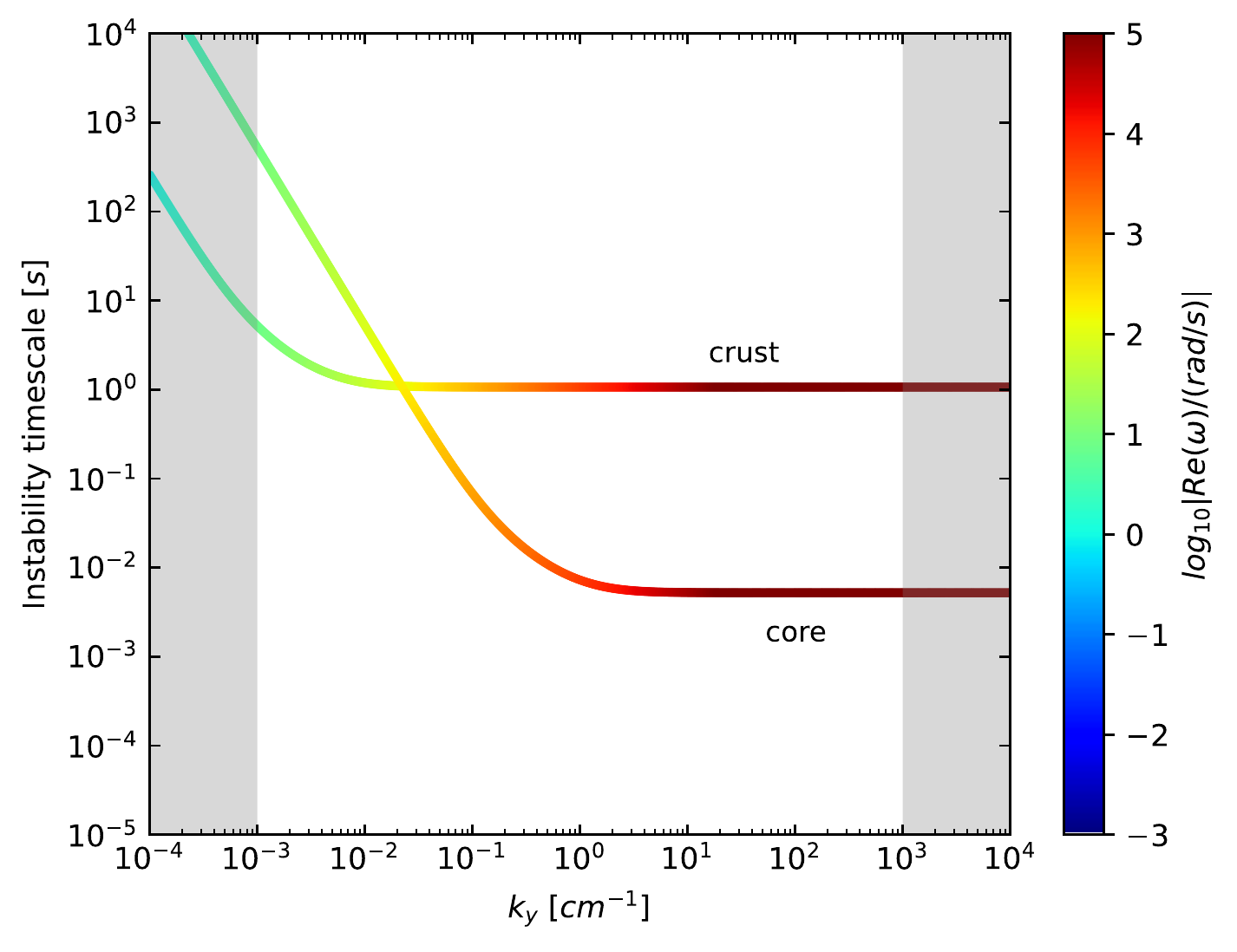}
\vspace{4ex}
\end{minipage} 

\caption{Instability timescale versus wave-vector $\mathbf{k}$ for the Two-Stream instability, in which the counterflow and $k$ are taken perpendicular to the vortex axis, in this case in the $y$ direction. The colour coding expresses the real part of the oscillation frequency (note that the rotation rate of the star, i.e. of the `normal' component, is taken to be $\Omega=100$ rad/s). As in previous cases, we consider the strong drag case with $\B=\B^{'}=0.5$ and two values for the background lag, a low value of $v_\n^z=1$ cm/s (left), and a high value (corresponding approximately to the maximum that pinning forces can sustain), of $v_\n^z=10^4$ cm/s (right). In general we see that there are always mixed inertial-sound waves, that are unstable on dynamical timescales, and that for large enough background velocity lags, the whole dynamical range is unstable.}
\label{Fig:2S3}
\end{figure*}

\begin{figure*}

\begin{minipage}[b]{0.49\linewidth}
\centering
\includegraphics[width=.95\linewidth]{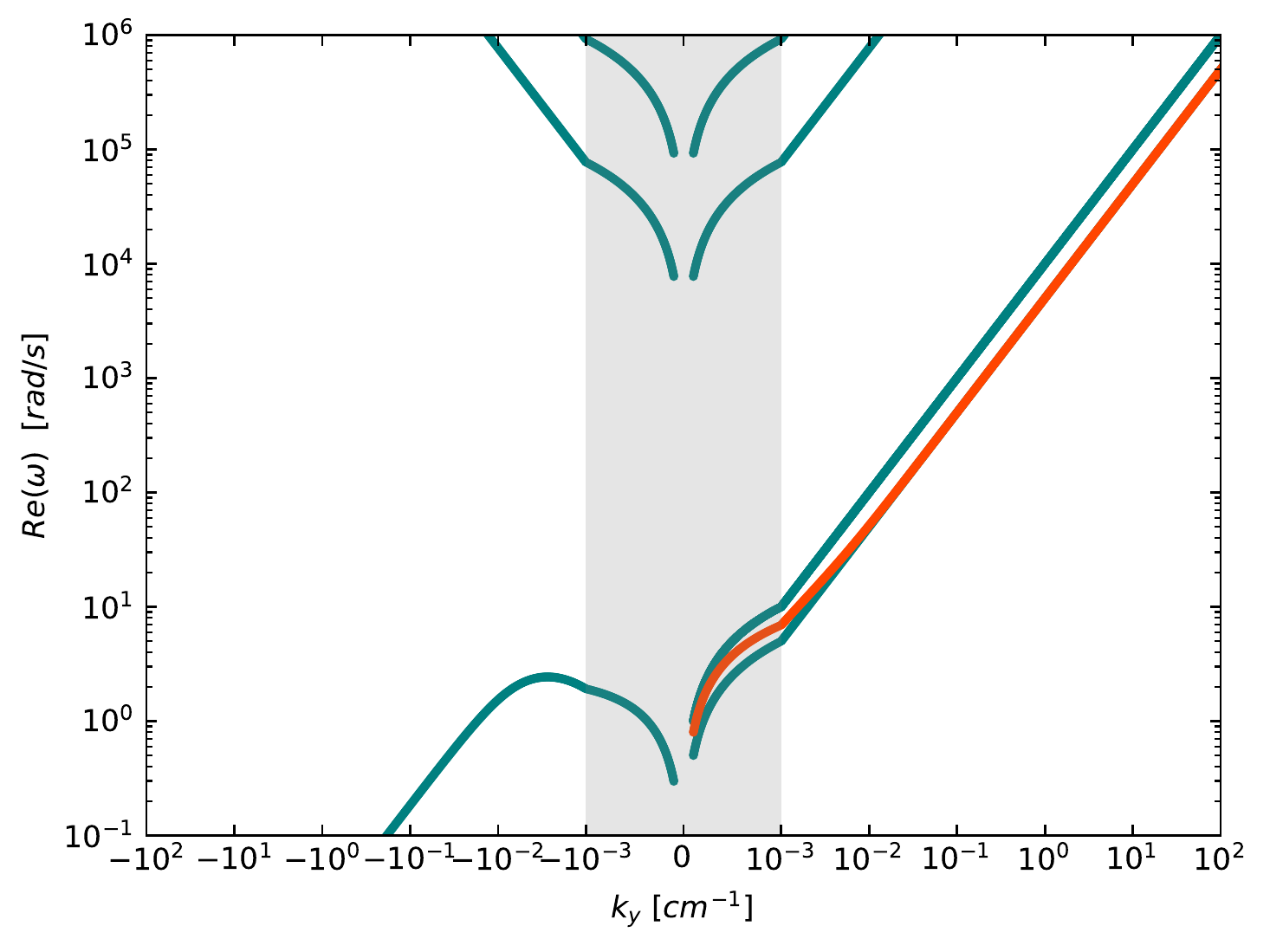} 
%\label{DG3}
%\label{fig:1vZkZ_1e0}
\vspace{4ex}
\end{minipage}
\begin{minipage}[b]{0.49\linewidth}
\centering
\includegraphics[width=.95\linewidth]{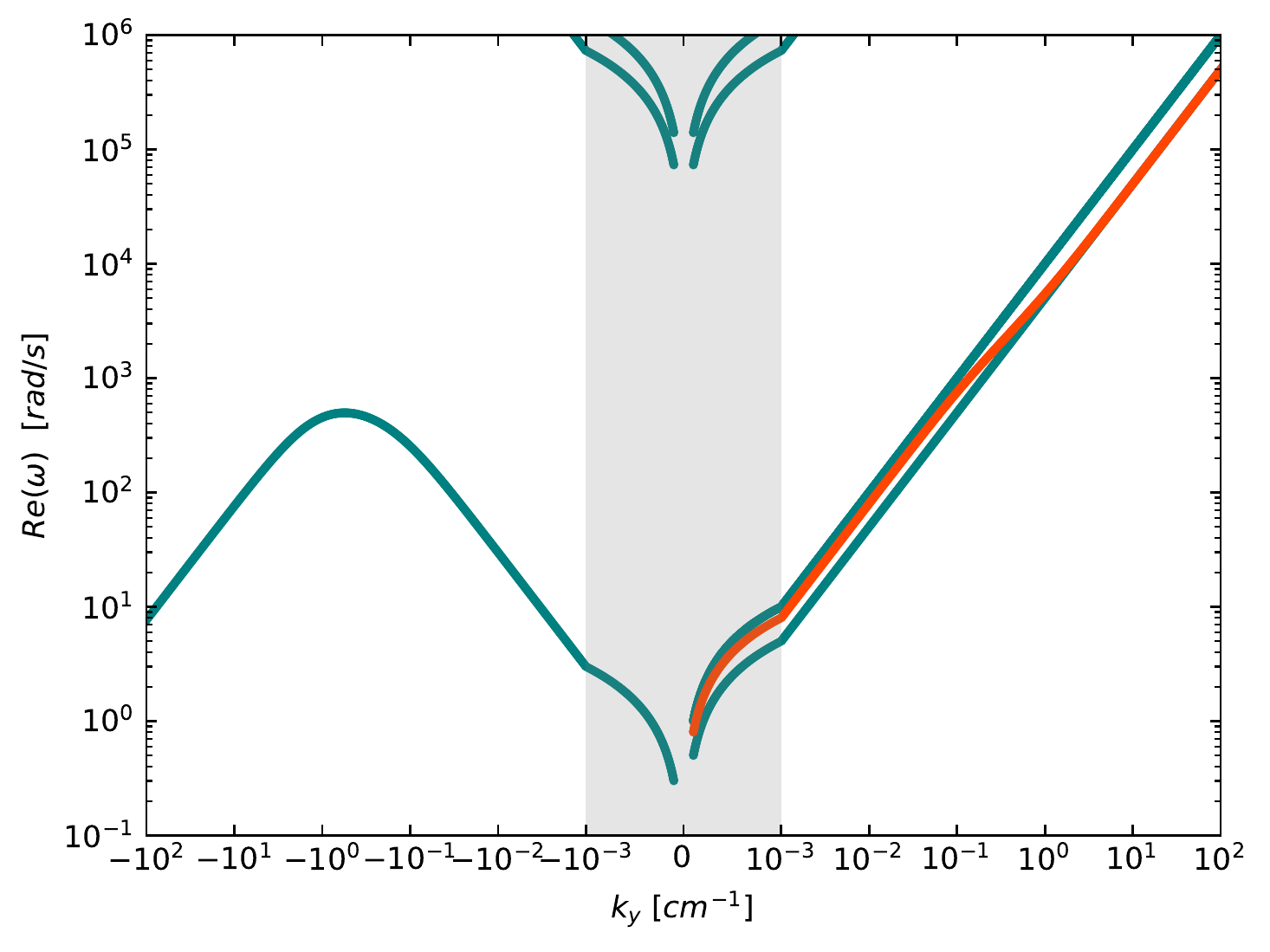}

%\label{fig:3vYkY_1e0}
\vspace{4ex}
\end{minipage} 

\caption{Dispersion relations (Real part of $\omega$ vs $k_y$) for the crust (left) and core (right) of the neutron star, for a lag of $v_\n^z=10^4$ cm/s for the strong drag case ($\mathcal{B}=\mathcal{B}^{'}=0.5$), for the Two-Stream instability in which we consider the background lag perpendicular to the vortex axis, i.e. the $y$ axis (the vortex is taken to be aligned with the $z$ axis). The unstable modes (in red) appear to be sound-inertial waves. Note that for the pinned or weak drag case, no instabilities are present. The shading identifies the region in which the scale of the horizontal axis is linear.}
\label{Fig:Disperdo3}
\end{figure*}

First of all we consider the DG instability, for strong and weak drag and in the `pinned' scenario, which we mimic by taking $\mathcal{B}^{'}=1$, but $\mathcal{B}=0$. In Figures \ref{Fig:DG3} and \ref{Fig:DG4} we plot the instability timescale for strong drag (weak drag is qualitatively similar) and for pinned vortices,  for varying values of the background lag.
In general we confirm the results of the analysis in section \ref{analyticDG}, there exists a family of mixed inertial and sound waves, that are unstable on dynamical timescales, both in the core and crust. Mutual friction has little effect on damping the instability (in fact it is at the heart of driving it), and rather it is the lag that plays a role in determining the onset of the instability. For small values of the lag of $v_\n^z=1$ cm/s, corresponding to $\Delta\Omega\approx 10^{-5}$ in the outer core or inner crust, assuming  a radius of $R\approx 10$ km, only the shortest length-scales are unstable. For higher lags ($v_\n^z=10^{4}$ cm/s, $\Delta\Omega\approx 10^{-2}$) all the dynamical range is unstable, and we can thus assume that once pinning allows for a significant enough lag to develop, the array will go unstable, possibly playing a role in triggering pulsar glitches or contributing to timing noise.
In Figures \ref{Fig:Disperdo1} and \ref{Fig:Disperdo2} we plot the dispersion relation for both the strong drag and pinned case. We see that the nature of the unstable modes changes: in the pinned case inertial modes are unstable, while in the strong drag case (and we have verified that the same is true for weak drag) the unstable mode is a  modified sound wave.

In Figure \ref{Fig:2S3} we consider the two-stream instability, in which the background counterflow is taken perpendicular to the vortex (and rotation) axis. Here again we confirm our analytical results from section \ref{AnalyticTS}. There are mixed sound-inertial waves (as can be verified from the dispersion relation in Figure \ref{Fig:Disperdo3}) that are unstable on dynamical timescales and the unstable range depends strongly on the magnitude of the background counterflow velocity. For high values, as may be expected from strong pinning, $v_\n^x=10^4$ cm/s, the whole dynamical range is unstable, with the instability developing on similar timescales at all scales. For  $v_\n^x=1$ cm/s, on the other hand, only the smallest length-scales are unstable, and the instability develops on longer timescales, that may be affected by other viscous mechanisms, such as shear viscosity.
\section{Conclusions}

We have studied the modes of oscillation of a superfluid neutron star, accounting for background counterflows and including entrainment. We have considered the effect for both standard HVBK mutual friction that arises if the vortex array is straight, and the Gorter-Mellink isotropic form, relevant for fully developed turbulence. 

We find that for standard mutual friction there is always a fast instability in the case of counterflow along the vortex axis (which may arise, for example if the star is precessing, or if vortices bend on a large scale), which is the neutron star analogue of the Donnelly-Glaberson instability that is observed in laboratory studies of superfluid helium. 
We also confirm the existence of a rapid two-stream instability when the background counterflow is perpendicular to the vortex axis (as is expected if a difference in angular velocity between the superfluid and the normal fluid arises due to pinning).

We find that entrainment plays an important role in the instability, and its effect is twofold. On the one-side it extends the dynamical range over which inertial modes are unstable in the crust, and but on the other it significantly shortens the growth time in the core of the star.

When we consider isotropic Gorter-Mellink mutual friction we find that the previous instabilities are stabilized. This suggests that once a large enough lag sets in, the instabilities we have presented will disrupt the straight vortex array and a turbulent tangle will develop. At this point the system is stable until the turbulence decays and the process will start again. It is thus very likely that transitions to turbulence play a role in triggering pulsar glitches and in timing noise \citep{2006Peralta, ltiming}.

An important effect that we have not considered is that of the magnetic field. The problem is complicated, as the interior field configuration of a neutron star is generally unknown, but \citet{Link2018} found that for simple field geometries the two-stream instability can be stabilized, while the Donnelly-Glaberson instability is always present. We have mimicked the effect of the magnetic field by studying a setup in which the proton fluid is `clamped' and not affected by the mutual friction (to imitate the situation in which it is held in place by magnetic stresses) and find that the two-stream instability is still present in the weak drag regime when entrainment is present. Future work should focus on the full problem, accounting also for superconductivity in the core and the interaction between neutron vortices and superconducting flux tubes \citep{GAS}.

\section*{Acknowledgments}

We acknowledge support from the Polish National Science Centre grant SONATA BIS 2015/18/E/ST9/00577. Partial support comes from PHAROS, COST Action CA16214. We thank the Institute for Nuclear Theory at the University of Washington for its kind hospitality and stimulating research environment. This research was supported in part by the INT's U.S. Department of Energy grant No. DE-FG02- 00ER41132.

\bibliographystyle{mnras}
%\begin{thebibliography}{99}

\bibliography{Library2}
%\end{thebibliography}

\appendix
\section{Derivation of the dispersion matrix}
\label{appendixequations}

In this appendix we provide a more self-contained description of how to perform the local variations by using directly the plane wave ansatz and find the dispersion matrix of the two-fluid system.
In the following, the indexes $\x$ and $\y$ are always meant to be different ($\x \neq \y$), while $\x$ and $\y'$ can assume the same value.

For clarity, let us list all the  quantities that appear into the Euler equations \eqref{Euler} as the entries of a vector $Q$, namely
\beq
Q \, = \, (  
\,  v_\x^i , \, 
\rho_\x , \, \mux , \,  \epsx , \, p_\x^i , \, \omega^i  , \, \hat{\omega}^i  
\, ) \, .
\label{qlist}
\eeq
Note that the gravitational potential $\Phi$ is not included into the set of variables that we perturb, as well as the mutual friction parameters $\mathcal{B}$, $\mathcal{B}'$ and $A_{GM}$.
The angular velocity $\Omega$ is also kept constant and defines the rotating frame in which $\bf k$ and the pulsation $\omega$ of the plane wave ansatz \eqref{Qplane} are measured.  Since the quantities in $Q$ are not all independent, it is convenient to calculate their variation following the order in the list.
Once the quantities in $Q$ have been expressed in terms of the velocity fields only, the vector is expanded at the frist order as
\beq
Q\left[ 
\vb_\x + \bar{\vb}_\x \, e^{i\left(  \mathbf{k}\cdot  \mathbf{x} - \omega t \right)} \right]
\approx Q_B +  \sum_{\y} \, \dfrac{\partial Q}{\partial \bar{v}^j_\y} \,  \bar{v}^j_\y \, .
\label{qwert}
\eeq
Hence, the generic amplitudes $\bar{Q}$ introduced in equation \eqref{Qgeneral} are obtained by means of the coefficients
\beq
 Q^\y_j \, = \, e^{-i\left(  \mathbf{k}\cdot  \mathbf{x} - \omega t \right)} \, \dfrac{\partial Q}{\partial \bar{v}^j_\y} \bigg{|}_B 
 \, = \,  \dfrac{\partial \bar{Q}}{\partial \bar{v}^j_\y} \bigg{|}_B \, .
 \label{qwerty}
\eeq
For $\rho_x$, $ \mux$ and $\epsx$ the result has already been given in section \ref{planew}. From the contintuity equation we find the diagonal elements $\rho^{\x\x}_j$ of the three $2\times 2$ matrices $ \rho^{\x\y'}_j= \delta_{\x\y'} \rho^{\x\x}_j$, see equation \eqref{rho_amplitude}. Now, the thermodynamic relations of section \ref{termo} allow us to calcualte the explicit form of 
\beq
\bar{\mu}_\x  \, = \,\mu^{\x n}_a \, \bar{v}_n^a  \, + \,    \mu^{\x p}_a  \, \bar{v}_p^a  
\label{muu}
\eeq
and
\beq
\bar{\varepsilon}_\x \, = \, \varepsilon^{\x n}_a \, \bar{v}_n^a  \, 
+ \, \varepsilon^{\x p}_a  \, \bar{v}_p^a  \, \, ,
\label{espss}
\eeq
where the matrices $\mu^{\x\y'}_j$ and $\varepsilon^{\x\y'}_j$ have been given in equations \eqref{mubar} and \eqref{amp_entr} respectively. Using  the result for $\rho^{\x\y'}_j$, the final expressions read
\beq
\begin{split}
\mu^{\x\x}_i
&\, = \, 
  \dfrac{c_\x^2\, k^i}{\omega - \mathbf{k} \cdot \vb_\x} + \alpha_\x w_{\x\y}^i
\\
\mu^{\x\y}_i
&\, = \, 
\dfrac{C_\x \, k^i}{\omega - \mathbf{k} \cdot \vb_\y} - \alpha_\x w_{\x\y}^i
\label{amp_mu}
\end{split}
\eeq
for the specific momentum and 
\beq
\begin{split}
\varepsilon^{\x\x}_i
&\, = \, 
 (\alpha_\x-\epsx) \dfrac{  \, k^i}{\omega - \mathbf{k} \cdot \vb_\x}+A_\x {w}_{\x\y}^i   
\\
\varepsilon^{\x\y}_i
&\, = \, 
\dfrac{\alpha_\y \rho_\y}{\rho_x} \dfrac{ k^i}{\omega - \mathbf{k} \cdot \vb_\y }-A_\x {w}_{\x\y}^i  
\label{amp_entr2}
\end{split}
\eeq
for the entrainment parameters.
%Equations \eqref{rho_amplitude}, \eqref{mubar} and \eqref{amp_entr} allow us to express variations of the thermodynamic quantities in terms of the amplitudes of the velocity fields $\bar{\mathbf{v}}_\x$ and will therefore be used to reduce the full dynamical problem to the linear form in \eqref{general}.

At this point it is immediate to address also the variations of the momenta and of the vorticity.
To give a concrete example of how the linear combination \eqref{Qgeneral} looks like when applied to vectorial quantities, we start by formally writing  the momentum amplitude as
\beq
\bar{p}^i_\x  \, = \, p^{\x n}_{ia} \, \bar{v}_n^a  \, + \,    p^{\x p}_{ia}  \, \bar{v}_p^a  \, .
\label{pppp}
\eeq
Performing the  expansion \eqref{qwert} on the momenta defined in \eqref{Momentum} immediately gives
\beq
\bar{p}^i_\x  \, = \, (1-\epsx)\, \bar{v}_\x^i  \, + \, \epsx\, \bar{w}_\y^i + v_{\y \x}^i \,
 \bar{\varepsilon}_\x \, .
\label{ppppp}
\eeq 
The last term can be expanded  thanks to \eqref{espss}, and \eqref{qwert} tells us that 
\beq
\begin{split}
p^{\x\x}_{ia}
&\, = \, 
(1-\epsx)\,  \delta_{ia} + w^{\y\x}_i \, \varepsilon^{\x\x}_a
\\
p^{\x\y}_i
&\, = \, 
\epsx \,  \delta_{ia} + w^{\y\x}_i \, \varepsilon^{\x\y}_a
 \, .
\label{amp_p}
\end{split}
\eeq
When the   entrainment coefficients \eqref{amp_entr2} are inserted into the above formulas, we see that the momentum perturbations have a correction which depends on the coefficient $A_\x$. Not surprisingly, these terms are of the second order in the background velocity lag;  it is not particularly convenient to neglect them here, but from the numerical point of view these terms are expected to have little effect on the calculated dispersion relation.

The last ingredient that we need is to write the vorticity perturbation  in the same fashion of \eqref{Qgeneral}, namely
\beq
\bar{\omega}_i \, = \, {\omega}^n_{ia} \,  \bar{v}_n^a \, + \, {\omega}^p_{ia} \,  \bar{v}_p^a \, .
\label{omm}
\eeq
Starting from equation \eqref{dw_gen}, or applying directly \eqref{qwert} on the definition of vorticity together with \eqref{espss}, it is straightforward to check that
\beq
\begin{split}
-i \,  {\omega}^n_{ia} 
&\, = \, 
(1-\varepsilon_n) \, \epsilon_{ija}k^j +  \epsilon_{ijl}k^j v_{pn}^l \varepsilon^{nn}_a  
\\
-i \,  {\omega}^p_{ia} 
&\, = \, 
\varepsilon_n\, \epsilon_{ija} k^j +  \epsilon_{ijl} k^j w_{pn}^l \varepsilon^{np}_a  \, .
\end{split}
\label{vorty}
\eeq
Alternatively, the above result can be found directly from \eqref{ppppp} by observing that
$\bar{\omega}_i = i \, \epsilon_{iab} \, k^a \, \bar{p}_n^b $. Similarly, the variation of the vorticity versor can be easily obtained by means of the projector in equation \eqref{perp}.

\subsection{Left hand side of the Euler equations}

Let us define the right hand side of equations \eqref{Euler}  as
\begin{multline}
\mathcal{E}^\x_i  \, = \,  
(\partial_t +  \vxd^{j} \nabla_j) \pxu_i + \epsx w^{\x\y}_j \, \nabla_i \, \vxd^{j}+
\\
+ \nabla_i \tilde{\mu}_x + 2 \, \epsilon_{ijk} \Omega^j \vxd^{k} \, ,
\end{multline}
where  the gravitational potential $\Phi$ and the centrifugal term have been ignored: we are  interested in the variation $\delta \mathcal{E}_\x^i$ and these two terms only contribute to define the background configuration. 
We extend the notation \eqref{Qplane} to $\mathcal{E}^\x_i$, namely
\beq
\delta \mathcal{E}^\x_i \, = \, \bar{\mathcal{E}}^\x_i \, e^{i\left( \mathbf{k} \cdot \mathbf{x} - \omega t \right)} \, ,
\eeq
%
%Equations \eqref{qwert} and \eqref{qwerty} applied 
where
\beq
\bar{\mathcal{E}}^\x_i \, = \, 
 \mathcal{E}^{\x\x}_{ia} \, \bar{v}_\x^a \, 
+ \, 
\mathcal{E}^{\x\y}_{ia} \,  \bar{v}_\y^a
 \, .
 \label{amp_E}
\eeq
Using systematically the results obtained so far,  a small amount  of algebra gives
\begin{multline}
i \,  \mathcal{E}^{\x\x}_{ia} \, = \,
(\omega -\mathbf{k}\cdot \vbx ) \left[ (1-\epsx) \delta_{ia} + w^{\y\x}_{i} \varepsilon^{\x\x}_a \right] +
\\
 - k_i(\epsx w^{\y\x}_{a} + \mu^{\x\x}_a)+ 2 \, i \, \epsilon_{ija}\Omega^j
\end{multline}
and 
\beq
i \,  \mathcal{E}^{\x\y}_{ia} \, = \,
(\omega -\mathbf{k}\cdot \vbx ) \left[ \epsx \delta_{ia} + w^{\y\x}_{i} \varepsilon^{\x\y}_a \right] 
 - k_i \, \mu^{\x\y}_a 
\eeq
In a dynamical regime in which the mutual friction is zero (i.e. in the rather ideal situation in which the vortex lines can be considered free and no drag nor pinning interactions act upon them), this result is sufficient to build the matrix $M$ of equation \eqref{general}. More explicitly, the six equations of the full linear system \eqref{general} can be also be written as
\begin{align}
\begin{split}
 M^{nn}_{ia} \, \bar{v}_n^a \, + M^{np}_{ia} \, \bar{v}_p^a & \, = \, 0 \, 
\\
 M^{pn}_{ia} \, \bar{v}_n^a \, + M^{pp}_{ia} \, \bar{v}_p^a & \, = \, 0 \, ,
 \end{split}
\end{align}
where the indexes $i$ and $a$ run over the three spatial indexes (the summation over $a$ is understood).
Therefore, it should  be clear that each block of the full matrix $M$ can be obtained as
\beq
 M^{\x \y'}_{ia}  \, = \,  \mathcal{E}^{\x\,\y'}_{i\, a} \, ,
\label{Mfree}
\eeq
in the free vortex limit.

\subsection{Perturbing the mutual friction}

%Temporarily reintroducing the background configuration label, we have to calculate
%%
%\beq
%\mathbf{f}^n
%\, = \, 
%B' \, \boldsymbol{\omega} \times \mathbf{v}_{np}
%\, + \,  
%B \, \hat{\boldsymbol{\omega}} \times (  \boldsymbol{\omega} \times  \vb_{np} )\, .
%\eeq

In order to treat consistently the mutual friction term it is convenient to start from the case $\x=n$ in \eqref{eq1}, namely
\beq
\mathbf{f}^n
\, = \, 
B' \, \boldsymbol{\omega} \times \mathbf{v}_{np}
\, + \,  
B \, \hat{\boldsymbol{\omega}} \times (  \boldsymbol{\omega} \times  \vb_{np} )\, .
\label{dfn}
\eeq
Once the variation $\delta \mathbf{f}^n$ has been obtained, the reaction on the normal component will just be given by
\beq
\delta \mathbf{f}^p
\, = \, 
-\left( \dfrac{\delta \rho_n}{\rho_p} - \dfrac{\rho_n}{\rho_p^2} \delta\rho_p\right ) \mathbf{f}^n
\, - \,  
 \dfrac{ \rho_n}{\rho_p} \, \delta\mathbf{f}^n \, .
 \label{dfp}
\eeq
Note that this last equation is valid also for the Gorter-Mellink mutual friction \eqref{GM}.
In complete analogy with what has already been discussed for equation \eqref{amp_E}, we may write
\beq
\bar{{f}}^\x_i \, = \, 
 {f}^{\x n}_{ia} \, \bar{v}_n^a \, 
+ \, 
{f}^{\x p}_{ia} \,  \bar{v}_p^a
 \, .
 \label{amp_f}
\eeq
The calculation leading to an explicit form of the coefficients   ${f}^{\x\y'}_{ia}$   is laborious and the final expression is complex enough to be useless in an analytic approach. 
However,  it is possible to implement an exact procedure that can be solved by any software for symbolic calculus. First, the amplitude in \eqref{amp_f} can be formally obtained as
\beq
\bar{{f}}^\x_i \, = \, 
e^{-i\left( \mathbf{k} \cdot \mathbf{x} - \omega t \right)} \sum_{Q}  \bar{Q}
\dfrac{\partial}{\partial \bar{Q} } \bar{{f}}^\x_i[Q_B + e^{i\left( \mathbf{k} \cdot \mathbf{x} - \omega t \right)} \bar{Q}]  \, ,
 \label{amp_f}
\eeq
where the derivative is evaluated on the background configuration. Therefore, the matrices  in \eqref{amp_f} are computed as
\beq
{f}^{\x\y '}_{ij}\,   = \, \sum_{Q}  
\dfrac{\partial \bar{f}^\x_i}{\partial  \bar{Q}} 
\, \dfrac{\partial \bar{Q}}{\partial \bar{v}^j_{\y '}} \, \bigg{|}_B
\,   = \, \sum_{Q}  \,  Q^{\y '}_j \, 
\dfrac{\partial \bar{f}^\x_i}{\partial  \bar{Q}}  \, \bigg{|}_B
 \, .
 \label{qwerty}
\eeq
In this way, it is possible to find the complete matrix $M$ of equation \eqref{M66}: not surprisingly, its form is defined in terms of the four $3 \times 3$ blocks 
\beq
M^{\x \, \y'}_{i \, j} \, = \,  \mathcal{E}^{\x \, \y'}_{i\,j} \, -\, f^{\x \, \y'}_{i\,j} \, .
\label{Mfinal}
\eeq
The determinant of this  $6\times 6$  matrix is a huge rational function of complex coefficients, which numerator defines a high degree polynomial in $\omega$ and in the components of $\mathbf{k}$. The roots of this polynomial define the dispersion relation $\omega(\mathbf{k})$ of the oscillation modes of the two-fluid system. In section \ref{num_res} the dispersion relations relative to some physically interesting cases are studied numerically within the symplifing assumptions that $C_\x=\alpha_x=A_\x=0$. Notice that, according to \eqref{amp_entr2}, this does not imply that entrainment variations are null (as it has been assumed  in particular in \eqref{eqLHS} and \eqref{eqRHS}).

%We stress that differentiation is always intended with respect to the coordinates $\mathbf{x}$, so that, according to \eqref{QB}, the unperturbed quantities can be regarded as constants; this justifies the second equality in equation \eqref{dw}, in which the  gradients of the background quantities have been neglected\footnote{
%Moreover, since the analysis is local and the background configuration is stationary, it is  also true that $\delta (D_\x Q)  = D_\x (\delta Q)$, where the substantial derivative \eqref{Dx} contains the unperturbed velocity $\vb_\x$.
%}.
%
% 
%Following this procedure, the perturbed version of the Euler-like equations \eqref{Euler} can be used to define a linear system of six equations for the six independent amplitudes $\bar{v}_\x^i$ which has the form
% 
%... \eqref{general} 
% 
%where the multi-index $(\x i)$ runs over the six combinations 
%$\{ n\,x, \,n\,y, \,n\,z, \,p\,x, \,p\,y, \,p\,z\}$. 
%The only way to have non trivial solutions of the above equation is to impose that the determinant of $M$, which turns out to be an high degree polynomial in the components of the wave vector $\mathbf{k}$ and of the pulsation $\omega$) is zero. The numerical analysis of the roots of this complex determinant will be addressed in section ...  

\label{lastpage}
\end{document}